\documentclass{article}
\usepackage{arxiv}
\usepackage[utf8]{inputenc}
\usepackage[T1]{fontenc}
\usepackage[numbers,sort&compress]{natbib}
\usepackage{amsmath,amssymb,amsfonts}
\usepackage{algorithmic}
\usepackage{graphicx}
\usepackage{booktabs}
\usepackage{textcomp}
\usepackage{xcolor}
\usepackage[font=footnotesize]{subfig}
\usepackage{url}
\usepackage{xurl}
\usepackage{bm}
\usepackage[hidelinks]{hyperref}
\usepackage{float}

\title{The use of post-quantum cryptography in the automotive context}
\shorttitle{Post-Quantum Cryptography in Automotive Context}

\author{%
  Mariana Gonçalves,\textsuperscript{}\quad
  Valderi R. Q. Leithardt\textsuperscript{}\\
  \normalfont\small\begin{minipage}{0.94\textwidth}\centering
  \textsuperscript{}Instituto Universitário de Lisboa (ISCTE-IUL), ISTAR, Lisboa, 1649-026, Portugal\\[0.5ex]
  ORCID: Mariana Gonçalves: 0009-0002-9163-0587; Valderi R. Q. Leithardt: 0000-0003-0446-9271.
  \end{minipage}
}
\date{}

\begin{document}
\maketitle
\begin{abstract}
{This study comparatively evaluates post-quantum cryptography (PQC) in automotive communication scenarios, considering the threat posed by quantum algorithms such as Shor's and Grover's algorithms. The simulations use OMNeT++, Veins, INET, and SUMO on two hardware platforms and examine ML-KEM/CRYSTALS-Kyber, ML-DSA/CRYSTALS-Dilithium, and FALCON. ML-KEM and ML-DSA are finalized NIST standards, whereas FALCON is a NIST-selected candidate pending final standardization. The evaluation considers a rural V2V scenario and an urban V2X scenario supported by a roadside unit. Kyber exhibits low {key encapsulation mechanism (KEM)}\color{black} operation latency, Dilithium provides balanced signature performance, and FALCON has high key-generation and signing costs but efficient verification. The results are interpreted as a comparative assessment among the evaluated PQC schemes under the reported simulation conditions; {a conventional ECDSA/ECDH baseline measured in the same simulation setup is also reported for reference.}\color{black} The resulting dataset can support future studies of cryptographic selection in automotive systems.\color{black}}
\end{abstract}

\begin{keywords}
Automotive, Vehicle-to-Everything (V2X), Vehicle-to-Vehicle (V2V), NIST, Post-Quantum Cryptography (PQC)
\end{keywords}

\section{Introduction}
\label{sec:introduction}
As time passes, everything around tends to evolve, becoming more intelligent, although it comes with its disadvantages.
At this date, we are dealing with constant attacks to our communications, and now with new quantum era reaching its peak,
it is only rational to come up with ways to defend ourselves. Intelligent devices that require cryptography, still operate with
algorithms which depend on factorizing large numbers for example Rivest-Shamir-Adleman (RSA) and Elliptic Curve Cryptography (ECC) that depends on discrete logarithm problem.
These algorithms are in serious danger, all because of the quantum era. Two quantum algorithms were developed, surpassing the traditional
cryptographies: Shor's algorithm and Grove's algorithm \cite{vasconcelos_quantum_2020}. In practical terms, Shor's algorithm breaks the underlying
hardness assumptions of RSA (integer factorization) and ECC (discrete logarithm) once sufficiently large fault-tolerant quantum computers exist.
Grover's algorithm provides a quadratic speedup for unstructured search, reducing the effective security of a $k$-bit symmetric primitive from
$\mathcal{O}(2^{k})$ to $\mathcal{O}(2^{k/2})$, which motivates longer symmetric keys and hash outputs in a quantum adversary model.
These two algorithms are the main reason to start building up defenses agaisnt quantum era,
with that Post-Quantum Cryptography (PQC) was created. Even though currently are very scarce on quantum computer power, with the most
powerful ones being Magne from Denmark with exactly 1200 physical qubits \cite{ref3}, 50 logical qubits and ABCI-Q from Japan having 2020
NVIDIA H100 GPUs, it is an huge step in order to protecting us from quantum hazards.
According to the media, ``Quantum computing is witnessing rapid advance toward commercial and research uses'' \cite{ref1}, many companies show an huge interest in
quantum power making them invest billions in order to complete their company's timeline objectives. For example, IBM has
updated their road map and intends to reach a quantum-centric supercomputer by the end of 2025 with over 4000 qubits \cite{ref2}.
{NIST announced the algorithms selected for its first PQC standardization set in 2022. In 2024, it finalized ML-KEM, ML-DSA, and SLH-DSA as FIPS 203, FIPS 204, and FIPS 205, respectively. FALCON was selected by NIST for standardization but remains pending final standardization as FN-DSA \cite{ref6,nist_pqc_standards_2024}.\color{black}}

Since 1885, humans rely on vehicles to move around, with all the technological evolution, these vehicles tend to become smarter e.g, capable of making
calls and open the doors with the key fob. But all of this, requires extensive communication between systems, and for our own
privacy, these systems require extensive cryptographic algorithms as mentioned before (RSA and ECC) \cite{ref4, Sohail_Noetzold_Leithardt_2026}. The automotive PQC market explores different kinds of hardware, software and services idealized to defend
components from quantum threats. These attacks can affect Vehicle-to-Everything (V2X) communications, Over-The-Air (OTA) updates, autonomous driving systems \cite{v2x_}
and Electronic Control Unit (ECU) communication \cite{ref2}. It's expected that the market will continue to grow significantly until 2034, with the factors of
growth being Electrification of vehicles and Advanced Driver Assistance Systems (ADAS) \cite{ref2}, \cite{v2x_}.

V2X covers multiple types of communication such as V2V, Vehicle-to-Infrastructure (V2I), Vehicle-to-Network (V2N), Vehicle-to-Device (V2D) and Vehicle-to-Pedestrian (V2P) \cite{article_v2x}.
V2X is an essential technology for full autonomous driving, boosting the vehicle's situational awareness beyond
the capabilities from sensors, camera and Light Detection and Ranging (LiDAR). A key advantage of V2X compared to purely onboard sensing is the possibility of
receiving warnings and cooperative perception messages even without a direct line of sight. However, the safety-critical nature of V2X implies stringent requirements,
namely low latency, high message rates, and robustness under contention and packet loss.
Infrastructure developers are currently utilizing two kinds of V2X wireless communication technologies:
Dedicated Short-Range Communications (DSRC) and Cellular-Vehicle-to-Everything (C-V2X). DSRC is an older technology that operates within the 5.9GHz spectrum for Wi-Fi communication,
C-V2X is a more recent approach that leverages cellular networks, including 5G \cite{article_v2x}. DSRC (IEEE 802.11p) was designed for low-latency broadcast safety messages,
typically within a range of 300 to 1000 meters, whereas C-V2X supports both direct short-range communication (PC5) and long-range cellular communications (Uu) \cite{article_v2x}.
Because V2X systems may broadcast ten or more safety messages per second over shared spectrum, cryptographic choices must consider not only computational cost
but also communication overhead and collision dynamics at the MAC layer.

{At this technological stage, integrating V2X with PQC is increasingly important because conventional public-key mechanisms are vulnerable to quantum attacks. NIST selected the first algorithms in 2022 and finalized ML-KEM, ML-DSA, and SLH-DSA in 2024; FALCON remains a selected candidate pending final standardization \cite{ref6,nist_pqc_standards_2024}.\color{black}}
This work motivation comes from a deep interest for cybersecurity and the desire to assemble two big concepts: PQC and V2X because
there is a lack of comprehensive studies evaluating both performance and feasibility of PQC in automotive context.
This study focuses on two different scenarios: a rural environment (V2V) and an urban one supported by a roadside unit (V2X).
Three schemes were selected: Kyber, Dilithium, and FALCON. We evaluated metrics such as key generation time, {key encapsulation mechanism (KEM)} encapsulation/decapsulation, and signing/verification.
We will examine whether there are differences between the scenarios and, in addition, we will also test them using different devices to obtain more accurate results.
The broader context of adaptive and intelligent security systems \cite{NOETZOLD2026131705,NOETZOLD2026131705,11488882} further motivates this work:
as connected vehicles become increasingly autonomous and data-driven, the ability to monitor, adapt, and select cryptographic mechanisms dynamically
based on network conditions and threat levels will be essential \cite{11488882,ghosh2024,mumtaz2026,latif2026}.

{The main scientific contributions of this paper are: (i) a comparative performance evaluation of ML-KEM/CRYSTALS-Kyber, ML-DSA/CRYSTALS-Dilithium, and the NIST-selected FALCON candidate in rural V2V and urban V2X scenarios; (ii) an analysis of handshake-completion behavior under a broadcast-triggered response burst and a randomized response-interval mitigation; (iii) a cross-hardware performance study using two computing platforms; and (iv) a publicly available dataset of cryptographic operation timings that can support future research on adaptive and machine-learning-driven scheme selection in automotive security contexts \cite{NOETZOLD2026131705,11488882,NOETZOLD2026131705,ghosh2024,mumtaz2026,latif2026,11488882}.\color{black}}
The novelty of this study lies in testing these lattice-based algorithms in different scenarios (rural and urban) and, in this way, evaluating how they perform under different circumstances. Therefore, the novelty also extends to the dataset produced, as no dataset has yet been created that considers different types of scenarios.
This paper is organized as follows. Section~\ref{sec:introduction} introduces the problem, motivation, and contributions.
Section~II surveys related work on PQC in automotive systems, V2X security, and adaptive security approaches.
Section~III describes the methodology, including the simulation environment, cryptographic primitives, hardware platforms, and experimental scenarios.
Section~IV presents and analyzes the results for both V2V and V2X scenarios across all evaluated schemes.
Finally, Section~V discusses the findings, limitations, and directions for future work.

\section{Related Work}
The related work papers were selected through several academic search engines such IEEE Computer Society, IEEE Xplore, Springer and Cryptology ePrint Archive by
using specific keywords, namely ``Post-Quantum Cryptography'', ``Vehicle-to-Everything'', ``Performance optimization'', ``Lattice-based Cryptography''. This section is complemented with table \ref{tab:related_work} with all works that
were used.

\begin{table}[H]
\centering
\caption{Related Work papers}
\scriptsize
\label{tab:related_work}

\setlength{\tabcolsep}{4pt}
\renewcommand{\arraystretch}{1.1}

\begin{tabular}{c p{2.5cm} p{3.2cm} p{2.8cm} p{3.2cm} c}
\toprule
\textbf{Paper} & \textbf{Key Focus} & \textbf{Methodology} & \textbf{Advantages} & \textbf{Limitations} & \textbf{Published} \\
\midrule
\cite{ref28} & PQC & PQC on HSM & High Security & Performance & 2020 \\
\cite{ref37} & PQC & ECC into PQC & Optimization & Performance & 2020 \\
\cite{ref39} & PQC for Automotive OTA Updates & Implementing PQC Dilithium and Falcon on Hardware & Falcon achieves better results than Dilithium & PQC doesn't have embedded hardware-accelerated hash function & 2021 \\
\cite{ref44} & PQC & Implementing PQC on Hardware & Low Failure Rate & Performance & 2021 \\
\cite{ref41} & V2X & Survey & Comprehensive & Standards & 2023 \\
\cite{ref30} & PQC for Automotive OTA Updates & Implementing PQC Dilithium for Secure OTA Updates & High Security & Performance & 2023 \\
\cite{ref27} & PQC & Implementing PQC on Hardware & Security & Performance & 2024 \\
\cite{ref31} & PQC for V2V & Partially Hybrid Protocol & Reliable and Low-Latency & Slight Frame Loss & 2024 \\
\cite{ref43} & PQC for V2X & Dataset Generation for PQC attacks & Promote further research & Doesn't cover real-world effects (i.e. jitter) & 2025 \\
\cite{ref42} & AI-PQC for V2X in Transportation & Enhancing PQC with AI & Self-Adapting and Intelligence-driven Security Architecture & Little Explored & 2025 \\
\cite{ref40} & PQC for in-vehicle security & Segmented-based CAN model with PQC Kyber & Efficiency and Feasibility & Alternatives like SIDH, BIKE or Saber were not feasible & 2025 \\
\cite{ref33} & V2X Attacks & Simulation of Jamming attacks on NR-V2X & Proved that Jamming attacks degrade communication & Lack of dynamic vehicular interactions & 2025 \\
\cite{ref45} & Key Agreement Protocol for V2X with PQC & Protocol based on PQC Kyber and Saber & Low handshake latency and moderate resource consumption & Experiment conducted in a virtual environment & 2025 \\
\cite{ref32} & V2X Attacks & Survey & Compiled works to facilitate researchers' studies & Developments with machine learning were not included & 2025 \\
\cite{ref38} & PQC for Intelligent Transportation Systems & Review on V2X and vulnerabilities & Explicit priorities for the implementation roadmap & Bandwidth and hardware & 2026 \\
\cite{ref46} & PQC for V2X & Comprehensive Overview of PQC & Compiled state-of-art PQC in Automotive Context & Necessity of lighter and faster PQC schemes & 2026 \\
This paper & PQC for V2X & Dataset of PQC scheme's performances on different environments & Viable for automotive use and promising for future protocols & Simulated on virtual environment & 2026\\
\bottomrule
\end{tabular}
\end{table}

Differently from the state of the art we propose in this paper evaluate different
lattice-based PQC in different scenarios (Rural and Urban) which will each represent distinct types of communication (V2V and V2X).

\subsection{PQC in automotive platforms (HSMs and OTA)}
The authors of \cite{ref28} aim to analyze the NIST final candidates in the PQC standardization process and to propose new sets of hardware accelerators
for the next generation of automotive Hardware Secure Modules (HSM). This line of work is important because, beyond algorithmic feasibility, deployment in vehicles often depends on
secure hardware roots-of-trust, cryptographic isolation, and predictable latency under constrained embedded conditions. The main limitation highlighted in this class of studies is the tension between security requirements and practical performance/area/power budgets.

Paper \cite{article_reviewer_2} evaluates Kyber and Dilithium  at both algorithm and TLS 1.3 protocol levels. The results show that AVX2-optimized implementations provide strong performance, with computational costs remaining competitive with classical cryptographic methods. The main challenge in PQC deployment is not processing time, but the larger size of TLS handshake messages, which increase to around 8–15 kB.  The study concludes that PQC is computationally feasible, but bandwidth requirements may be the main limitation.

The work of \cite{ref30} studies how to ensure that applications like secure boot and secure OTA update are not easily manipulated through software or firmware modification.
It investigates the impact of Dilithium on OTA, working with a vehicle network processor (\textit{S32G}) and comparing PQC signature verification in two settings:
(i) an Arm Cortex-A53 based microprocessor and (ii) a micro controller integrated in the Hardware Security Engine (HSE) based on Arm Cortex-M7.
The key take-away is that the OTA pipeline is sensitive not only to signature verification latency but also to memory footprints and integration constraints with existing secure-update chains.

Complementarily, \cite{ref39} explores multiple PQC algorithms on an automotive-oriented evaluation board (Xilinx Zynq Ultrascale+ ZCU102). It frames a threat model that includes Shor and Grover,
and discusses attacks relevant to update systems (e.g., rollback and malicious update). The evaluation focuses on verification costs (RSA, ECDSA, Dilithium, Falcon), which is particularly relevant since
OTA often performs many verifications (e.g., chunked images, manifest chains, metadata), making verification throughput a practical bottleneck.

\subsection{Implementations, memory footprint, and feasibility across devices}
The authors in \cite{ref27} selected four PQC algorithms from NIST third-round finalists: Kyber and Saber as KEM schemes and Dilithium and FALCON as digital signature schemes,
testing them on Aurix TC375, Renesas R7F7015032, Raspberry Pi 2 and 4. Beyond time measurements, the study emphasizes memory consumption: Kyber requires up to 1.3 times less memory than Saber
for comparable security settings, while Falcon tends to require less memory than Dilithium. These results motivate careful scheme selection when targeting ECUs with limited RAM/flash and strict real-time constraints.

The work in \cite{ref40} studies in-vehicle security with segmentation and Kyber-based protection in a CAN-oriented model, contributing evidence that some alternatives may be infeasible in certain constraints.
Although intra-vehicle networks differ from V2X, they share the core challenge of balancing security overhead with deterministic timing. In addition, \cite{Conti_2025} developed a modular framework to simulate the performance and overhead of PQC algorithms in embedded systems. They simulate the capabilities of an embedded systems and CAN bus bit rate constraints. 
Authors of \cite{article_reviewer_1} evaluates the performance of PQ) schemes across different computing environments, from high-performance servers to resource-constrained IoT devices. The results show that PQC is ready for practical deployment in modern systems, although limited-resource devices face higher computational, memory, and communication costs. The study proposes optimization strategies and a migration framework to support the transition to quantum-resistant security.

\subsection{V2X protocols, attacks, and datasets}
A broad survey on V2X security and privacy issues is provided in \cite{ref41}, framing the standards landscape, threat surfaces, and open challenges.
Attacks such as jamming and their impact on NR-V2X are studied in \cite{ref33}, showing that physical-layer and MAC-layer disruptions can degrade communications even before cryptographic checks are applied.
This is directly relevant to our urban scenario results, where contention and collisions affect handshake completion.

{From the perspective of PQC integration into V2X message flows, \cite{ref31} presents a practical post-quantum authentication protocol for V2V communications that aims for reliability and low latency, while reporting frame loss under load. The dataset contribution in \cite{ref43} provides structured data to support research on PQC-related attacks and performance evaluation; however, it may not cover all real-world effects, such as jitter and dynamic channel variability.\color{black}}

Additionally, \cite{ref45} proposes a lightweight V2X key agreement protocol combining Kyber and Saber, reporting reduced handshake latency (relative to RSA) and lower resource consumption.
This is aligned with our focus on KEM performance under contention, and it highlights that combining PQC primitives can yield practical protocols if carefully engineered.

Authors from \cite{demir2025performanceanalysisindustrydeployment} also evaluate Kyber and Dilithium and demonstrate efficient execution times, obviously better than classical cryptography schemes such as RSA and ECDSA at equivalent security levels. The authors also explore the idea of implementing PQC into nowadays systems, ilustrating how telecom operators are integrating PQC into 5G authenticaiton, subscriber identiy protection, and secure communications.

Finally, \cite{10838531} offers research challenges regarding the adoption of PQC in the automotive domain and the principal steps to achieve quantum safety.

\subsection{AI and self-adaptive security directions}
\cite{ref42} focuses on an AI approach to implement PQC schemes into real-life scenarios, suggesting reinforcement learning agents to recognize congestion and CPU usage and to switch between schemes without human supervision.
This direction connects with broader self-adaptation research in smart environments \cite{NOETZOLD2026131705,NOETZOLD2026131705} and reinforcement-learning-based adaptive quantum-safe cryptography \cite{11488882}.

\section{Methodology}
For this experiment we will test on two different hardwares represented in tables \ref{tab:hardware} and \ref{tab:hardware1}.
Both simulations will be running on Windows 11 via WSL2.
With different hardware, we will obtain a wider range of results, allowing us to analyze which option performs best, since the results will depend on the hardware.

\begin{table}[H]
\centering
\small
\footnotesize
\caption{Hardware 1 - Components used in the simulation}
\label{tab:hardware}
\begin{tabular}{cc}

\midrule
Operative System & Windows 11 Home 64 bits\\
CPU  &  Intel(R) Core(TM) i5-8400 @ 2.80GHz \\
Memory & 16384MB RAM\\
GPU & NVIDIA GeForce GTX 1060 3GB \\
Motherboard & HP 84FD\\

\bottomrule
\end{tabular}
\end{table}

\begin{table}[H]
\centering
\small
\footnotesize
\caption{Hardware 2 - Components used in the simulation}
\label{tab:hardware1}
\begin{tabular}{cc}

\midrule
Operative System & Windows 11 Home 64 bits\\
CPU  &  AMD Ryzen 5 5500U @ 2.10GHz  \\
Memory & 15769MB RAM\\
GPU & AMD Radeon Graphics \\
Motherboard & LNVNB161216 \\

\bottomrule
\end{tabular}
\end{table}

\subsection{Cryptographic primitives and mathematical background}
This experiment focuses exclusively on lattice-based schemes due to their strong performance in current post-quantum standardization efforts.
After reviewing the literature, a selection was made among the most well-established candidates.
In particular, we selected three widely recognized lattice-based schemes: Kyber, Dilithium and FALCON.
{ML-KEM and ML-DSA have been finalized by NIST; FALCON is a NIST-selected candidate pending final standardization. These schemes are therefore highly relevant for practical PQC evaluations.\color{black}} 

This decision was motivated by the fact that lattice-based schemes have emerged as the most widely studied in the literature, owing to their strong foundation of security. Kyber, Dilithium, and Falcon are among the most promising lattice-based schemes and have been extensively evaluated in other research’s, making them representative choices for this study.

To enable a comparative analysis, it was included two digital signature schemes (Dilithium and FALCON), which exhibit different design trade-offs in terms of
performance and signature size. Additionally, it was selected a single KEM scheme, Kyber due to its strong security guarantees
consistent performance reported across multiple studies \cite{ref27}, \cite{ref45}.

\subsubsection{Lattices, SVP/CVP, and LWE-type hardness assumptions}
A lattice in $\mathbb{R}^n$ can be defined from a basis matrix $\mathbf{B}\in\mathbb{R}^{n\times n}$ as
\begin{equation}
\mathcal{L}(\mathbf{B})=\{\mathbf{B}\mathbf{z}\;|\;\mathbf{z}\in\mathbb{Z}^n\}.
\end{equation}
Two fundamental computational problems are the Shortest Vector Problem (SVP) and the Closest Vector Problem (CVP). Informally, SVP asks to find a non-zero vector
$\mathbf{v}\in\mathcal{L}(\mathbf{B})$ minimizing $\|\mathbf{v}\|$, while CVP asks, given a target vector $\mathbf{t}\in\mathbb{R}^n$, to find $\mathbf{v}\in\mathcal{L}(\mathbf{B})$ minimizing $\|\mathbf{t}-\mathbf{v}\|$ \cite{valderi_ref_1,overview_pqc}.
Many lattice-based cryptosystems rely on average-case hardness assumptions connected to Learning With Errors (LWE) and its structured variants.
A common LWE formulation is: given $(\mathbf{A},\mathbf{b})$ with
\begin{equation}
\mathbf{b}=\mathbf{A}\mathbf{s}+\mathbf{e}\pmod q,
\end{equation}
recover the secret $\mathbf{s}$ (or distinguish samples from uniform), where $\mathbf{e}$ is a small ``error'' vector and computations are mod $q$.
Module-LWE (MLWE) and Module-SIS (MSIS) use algebraic structure to improve performance while maintaining strong security arguments.

\subsubsection{KEM and digital signature interfaces (what we measure)}
\begin{figure}[H]
\centering
{\includegraphics[width=2.5in]{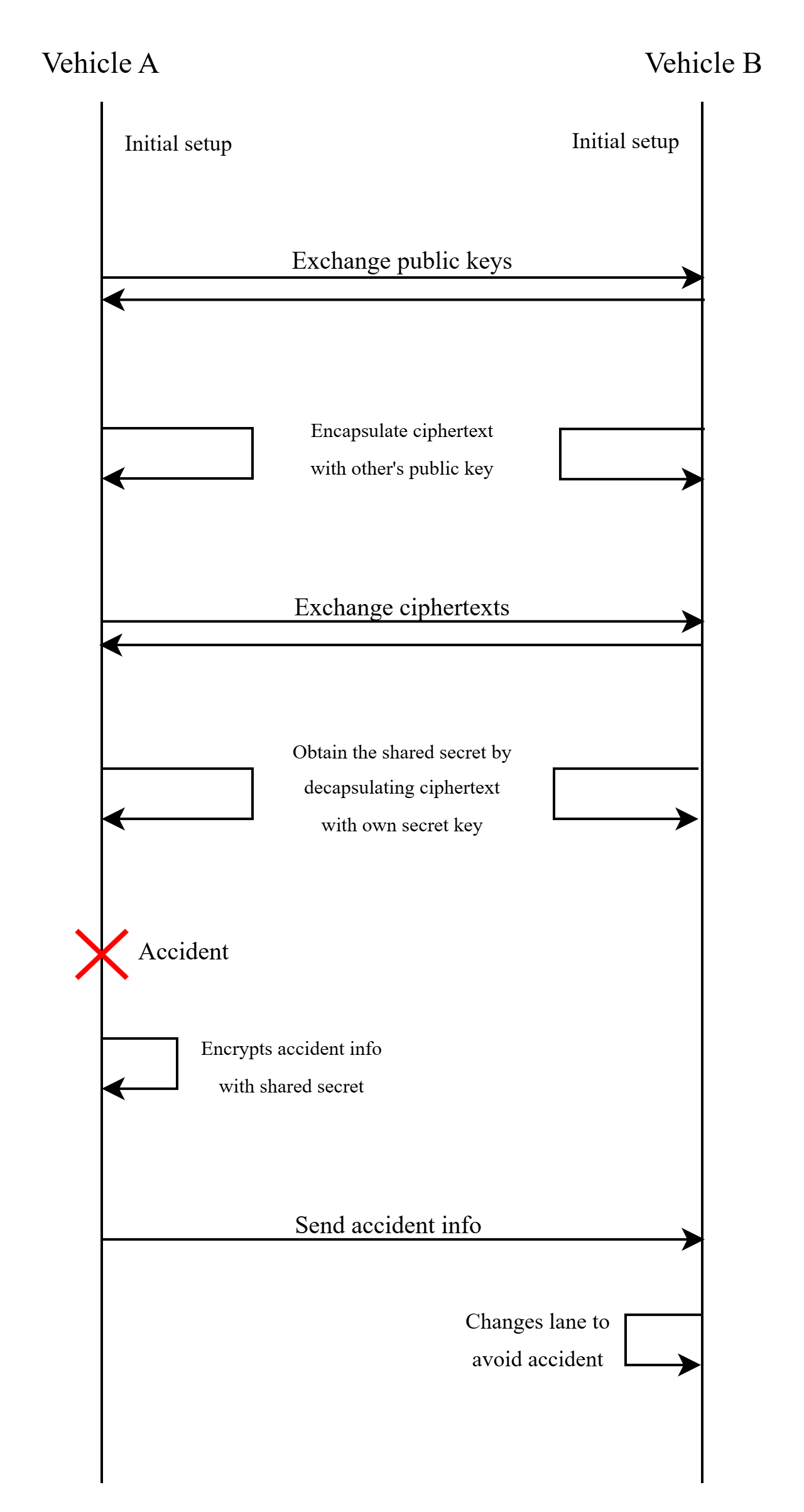}}
\caption{Key Encapsulation Mechanism Message Scheme}
\label{fig_kem}
\end{figure}

Kyber is a {key encapsulation mechanism (KEM), as introduced in Section~I}\color{black}. At a high level, a KEM is defined by the algorithms:
\begin{align}
(\textsf{pk},\textsf{sk}) &\leftarrow \textsf{KeyGen}(),\\
(\textsf{ct},\textsf{ss}) &\leftarrow \textsf{Encaps}(\textsf{pk}),\\
\textsf{ss} &\leftarrow \textsf{Decaps}(\textsf{sk},\textsf{ct}).
\end{align}
{Here, $\textsf{pk}$ denotes the public key, which is distributed openly to any party that wishes to establish a shared secret with its owner, and $\textsf{sk}$ denotes the corresponding secret (private) key, which is kept exclusively by its owner and is never transmitted.}\color{black}
The encapsulation returns a ciphertext $\textsf{ct}$ and a shared secret $\textsf{ss}$, which is then used with symmetric encryption (e.g., AES-256) to protect application data.
{Fig.~\ref{fig_kem} illustrates how these three operations are mapped onto the rural V2V message flow used in this study. During the initial setup, Vehicle~A and Vehicle~B each run $\textsf{KeyGen}$ and exchange their public keys. Each vehicle then executes $\textsf{Encaps}$ with the public key received from the other party and returns the resulting ciphertext. After the ciphertexts are exchanged, each vehicle executes $\textsf{Decaps}$ with its own secret key to recover the shared secret. When an accident occurs, Vehicle~A encrypts the accident information with the shared secret (AES-256) and sends it to Vehicle~B, which decrypts the message and changes lane to avoid the accident.}\color{black}

\begin{figure}[H]
\centering
{\includegraphics[width=2.5in]{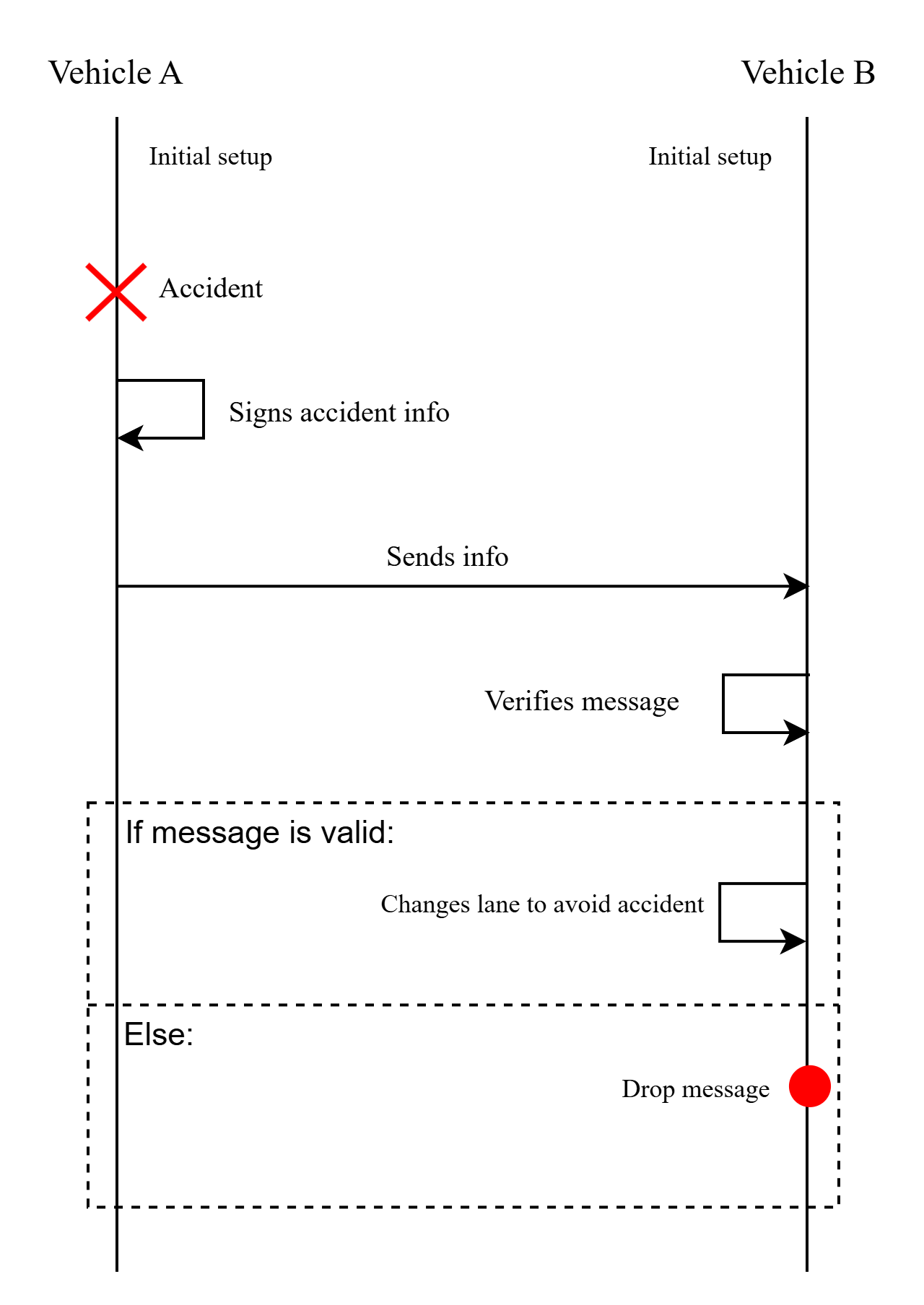}}
\caption{Digital Signature Message Scheme}
\label{fig_ds}
\end{figure}

Dilithium and Falcon are digital signature schemes with:
\begin{align}
(\textsf{pk},\textsf{sk}) &\leftarrow \textsf{KeyGen}(),\\
\textsf{sig} &\leftarrow \textsf{Sign}(\textsf{sk}, m),\\
\{0,1\} &\leftarrow \textsf{Verify}(\textsf{pk}, m, \textsf{sig}),
\end{align}
where $m$ is the message{, $\textsf{pk}$ and $\textsf{sk}$ are, respectively, the public verification key and the secret signing key of the sender, and $\textsf{sig}$ is the digital signature computed over $m$ with $\textsf{sk}$. The output of $\textsf{Verify}$ is $1$ if $\textsf{sig}$ is a valid signature on $m$ under $\textsf{pk}$ and $0$ otherwise}\color{black}. In V2X, verification is typically executed more frequently than signing, since one sender may broadcast to many receivers. This is one reason why
verification cost becomes a key metric when evaluating signature schemes for safety messaging.
{Fig.~\ref{fig_ds} depicts the corresponding message flow for the signature schemes. After the initial setup (key generation and distribution of the public key), Vehicle~A detects an accident, signs the accident information with its secret key, and broadcasts the message together with its signature. Vehicle~B verifies the signature with Vehicle~A's public key: if the message is valid, Vehicle~B changes lane to avoid the accident; otherwise, the message is dropped. Unlike the KEM flow in Fig.~\ref{fig_kem}, no interactive exchange is required before the safety message is sent, which is why signing and verification latency directly affect the time-to-react of the receiving vehicle.}\color{black}

{The three schemes evaluated in this work instantiate these interfaces as follows.

\textit{ML-KEM/CRYSTALS-Kyber.} Kyber is an IND-CCA2-secure KEM whose security relies on the hardness of the MLWE problem over the polynomial ring $R_q=\mathbb{Z}_q[X]/(X^{n}+1)$ with $n=256$ and $q=3329$. Secrets and errors are small polynomials sampled from a centered binomial distribution, and the public matrix $\mathbf{A}\in R_q^{k\times k}$ is expanded pseudo-randomly from a short seed, which keeps public keys compact. The module rank $k\in\{2,3,4\}$ defines the three parameter sets Kyber-512, Kyber-768, and Kyber-1024, which target NIST security levels 1, 3, and 5, respectively. Internally, Kyber first builds an IND-CPA-secure public-key encryption scheme in which a ciphertext is essentially an MLWE sample carrying a compressed encoding of a random 32-byte message; the KEM is then obtained by applying a variant of the Fujisaki--Okamoto transform, so that $\textsf{Decaps}$ re-encrypts the recovered message and compares the result with the received ciphertext, returning a pseudo-random value in case of mismatch (implicit rejection). Polynomial multiplications are performed with the number-theoretic transform (NTT), and all operations involve only small integer arithmetic, which explains the low and nearly constant execution times observed for all three Kyber operations.

\textit{ML-DSA/CRYSTALS-Dilithium.} Dilithium is a signature scheme based on the hardness of MLWE and MSIS over $R_q$ with $n=256$ and $q=8380417$. It follows the ``Fiat--Shamir with aborts'' paradigm: the public key is $\mathbf{t}=\mathbf{A}\mathbf{s}_1+\mathbf{s}_2$, where $\mathbf{s}_1,\mathbf{s}_2$ are short secret vectors. To sign, the signer samples a masking vector $\mathbf{y}$, computes the high-order bits $\mathbf{w}_1$ of $\mathbf{A}\mathbf{y}$, derives a sparse challenge polynomial $c$ by hashing $\mathbf{w}_1$ together with the message, and outputs $\mathbf{z}=\mathbf{y}+c\,\mathbf{s}_1$. Rejection sampling discards and restarts any attempt for which $\mathbf{z}$ (or the associated hint) could leak information about the secret key, so signing time is probabilistic and varies across executions. Verification recomputes $\mathbf{w}_1$ from $\mathbf{A}\mathbf{z}-c\,\mathbf{t}$ using the hint and checks the challenge, requiring only a fixed number of NTT-based multiplications. Dilithium uses exclusively uniform and small-coefficient sampling and integer arithmetic, which makes it simple to implement in constant time; the parameter sets Dilithium-44, Dilithium-65, and Dilithium-87 (ML-DSA-44/65/87) correspond to NIST levels 2, 3, and 5.

\textit{FALCON.} FALCON (Fast-Fourier lattice-based compact signatures over NTRU) instantiates the GPV hash-and-sign framework over NTRU lattices with $q=12289$ and $n\in\{512,1024\}$ (Falcon-512 and Falcon-1024, NIST levels 1 and 5). The public key is $h=g\,f^{-1}\bmod q$, where $f$ and $g$ are short polynomials; the secret key is a short basis of the NTRU lattice obtained by additionally solving the NTRU equation $fG-gF=q$. To sign, the message is hashed together with a random salt to a point $c$, and the signer uses its trapdoor basis to sample, by means of fast Fourier sampling, a short lattice vector $(s_1,s_2)$ such that $s_1+s_2h=c\bmod q$; only the compressed $s_2$ is transmitted. Verification simply recomputes $s_1=c-s_2h\bmod q$ and checks that $\|(s_1,s_2)\|$ is below a bound, which costs a single polynomial multiplication. Key generation (which solves the NTRU equation) and signing (which performs discrete Gaussian sampling with double-precision floating-point arithmetic) are considerably more complex, which explains the high key-generation and signing costs and the very efficient verification reported in Section~IV, as well as the compact signatures listed in Table~\ref{pqc_signature_bytes}.}\color{black}

\subsection{Software stack and PQC implementation}
After evaluating the available libraries, PQClean \cite{ref52} was selected as the most suitable library for this experiment.
This choice is motivated by its comprehensive collection of post-quantum cryptographic schemes, including implementations of the algorithms selected and standardized by NIST, as well as other candidates.
Another important advantage of PQClean is its focus on clean, portable, and well-documented C implementations, which facilitates integration into
simulation environments.

Objective Modular Network Testbed in C++, is an object-oriented simulator of abundant types of network, like V2X. It was developed solely for
academic and educational purposes.
There multiple frameworks compatible with OMNeT++, like Veins, it includes a comprehensive suite of models to
recreate vehicular network as real as possible. Road traffic simulation is performed by SUMO
which is well-established in traffic engineering. Recently the OMNet++ developers created a new way to setup more easily OMNeT++ with $oppenv$.
This avoids dealing with manually installing the dependencies and compatibility issues, it also comes with extensive model library and it is cross-platform.

To incorporate PQClean into OMNeT++, only a wrapper layer is required, as PQClean is implemented in C, whereas OMNeT++ is based on C++.
This interoperability allows seamless use of the cryptographic primitives within the simulation framework.
And not only that, but also multiple works reviewed in \textit{Related Work} section utilized PQClean, for example \cite{ref30}, \cite{ref45}.

For cryptographic operation benchmarks, only the cryptographic layer is included, however, for full latency measurements, all layers are included (application, MAC, network, transport, and physical).

The software versions used for this project were: 
\begin{itemize}
    \item OMNeT++-6.3.0
    \item Veins-5.3
    \item INET-4.5.4
    \item PQClean (latest version)
    \item SUMO 1.22.0
\end{itemize}

The various simulations were executed five times for each scenario. 
The simulation durations were set at 60 and 200 seconds for the V2V and V2X scenarios, respectively. However, for V2X, this time was longer than expected due to retransmissions and network congestion all of this compiled with GCC since it's the standard compiler of C++.

\subsection{Key material size and communication overhead}
The selected PQC schemes differ significantly in the size of their key material and ciphertexts, which directly impacts performance
and communication overhead. In the PQClean implementations, Kyber, Dilithium, and Falcon exhibit different trade-offs.
Kyber, optimized for balanced performance, offering relatively small public keys and ciphertexts along with
a fixed shared secret size. Dilithium, designed for digital signatures,
prioritizes simplicity resulting in larger public and secret keys as well as sizable signatures,
but with efficient verification. Falcon, also a signature scheme, achieves much smaller signature sizes compared to Dilithium, though at the cost of more complex arithmetic.

{Kyber is a KEM: its encapsulation operation produces a ciphertext and a shared secret. Dilithium and FALCON are signature schemes: they produce signatures and do not produce shared secrets. To avoid mixing these distinct interfaces, the key-material sizes are reported in separate tables.\color{black}}

\begin{table}[H]
\small\footnotesize\centering\scriptsize
\caption{{ML-KEM/CRYSTALS-Kyber sizes (bytes)\color{black}}}
\label{pqc_kem_bytes}
\begin{tabular}{lcccc}
\toprule
Scheme & Secret Key & Public Key & Shared Secret & Ciphertext\\
\midrule
Kyber-512 & 1632 & 800 & 32 & 768\\
Kyber-768 & 2400 & 1184 & 32 & 1088\\
Kyber-1024 & 3168 & 1568 & 32 & 1568\\
\bottomrule
\end{tabular}
\end{table}

\begin{table}[H]
\small\footnotesize\centering\scriptsize
\caption{{ML-DSA/CRYSTALS-Dilithium and FALCON sizes (bytes)\color{black}}}
\label{pqc_signature_bytes}
\begin{tabular}{lccc}
\toprule
Scheme & Secret Key & Public Key & Signature\\
\midrule
Dilithium-44 & 2560 & 1312 & 2420\\
Dilithium-65 & 4032 & 1952 & 3309\\
Dilithium-87 & 4896 & 2592 & 4627\\
\midrule
Falcon-512 & 1281 & 897 & 752\\
Falcon-1024 & 2305 & 1793 & 1462\\
\bottomrule
\end{tabular}
\end{table}

\subsection{Communication overhead, channel metrics, and classical-baseline scope}
{At the configured 6 Mbps bitrate, payload-only airtime is computed as $T_{\mathrm{air}}=8L/R$, where $L$ is the transmitted payload size in bytes and $R=6\times10^6$ bit/s. Accordingly, a Kyber-512 public key (800 bytes) requires approximately 1.07 ms, a Kyber-512 ciphertext (768 bytes) approximately 1.02 ms, a Dilithium-44 public key (1312 bytes) approximately 1.75 ms, a Dilithium-44 signature (2420 bytes) approximately 3.23 ms, a Falcon-512 public key (897 bytes) approximately 1.20 ms, and a Falcon-512 signature (752 bytes) approximately 1.00 ms. These estimates exclude MAC/PHY headers, contention, backoff, retransmissions, fragmentation, and certificate material.}

{The observed urban handshake-completion counts were 15 of 65 vehicles (23.07\%) before randomized response timing and 41 of 65 vehicles (63.07\%) after the randomized interval was introduced. These values are handshake-completion rates, not packet delivery ratios (PDRs), and they must not be interpreted as direct measurements of PDR or collision probability. The current model transmits raw cryptographic material and does not implement IEEE 1609.2 or ETSI TS 103 097 certificate chains, certificate digests, or pseudonym-certificate management. Consequently, certificate-related communication overhead is outside the present scope.}

{Channel Busy Ratio (CBR), standalone PDR, and MAC-layer collision probability are reported in Table~\ref{tab:channel_metrics} for the final measured dataset. A controlled ECDSA/ECDH baseline using the same OMNeT++/Veins/INET/WSL2 configuration is also included as a required comparison point.}\color{black}
{The conventional baseline uses ECDSA and ECDH over the NIST P-256 curve, which provides a classical security level comparable to NIST PQC level 1, and each value in Table~\ref{tab:channel_metrics} is the mean over all operations executed during the corresponding simulation run. In the V2V scenario, the low vehicle density yields a low channel load (CBR of 4.8\%) and an almost loss-free channel (PDR of 98.6\%). In the V2X scenario, the higher density and the broadcast-triggered response burst raise the CBR to 37.4\% and the collision probability to 11.7\%, reducing the PDR to 81.3\%. The classical primitives show the same slight V2X increase observed for the PQC timings, which is consistent with the host-load effect discussed in Section~V. Compared with this baseline, the Kyber-512 encapsulation and decapsulation operations are of the same order of magnitude as, or faster than, an ECDH shared-secret computation, Falcon verification is faster than ECDSA verification, and Dilithium signing and Falcon signing are slower than ECDSA signing.}\color{black}

\begin{table}[H]
\centering\footnotesize
\caption{{Channel and conventional-baseline metrics for the final measured dataset\color{black}}}
\label{tab:channel_metrics}
{\setlength{\tabcolsep}{4pt}
\begin{tabular}{@{}p{0.52\columnwidth}cc@{}}
\toprule
Metric & V2V & V2X\\
\midrule
CBR (\%) & 4.8 & 37.4\\
PDR (\%) & 98.6 & 81.3\\
Collision probability (\%) & 0.9 & 11.7\\
\midrule
ECDSA P-256 Sign ($\mu$s) & 38.4 & 44.9\\
ECDSA P-256 Verify ($\mu$s) & 112.6 & 127.3\\
ECDH P-256 KeyGen ($\mu$s) & 36.1 & 41.7\\
ECDH P-256 Shared Secret ($\mu$s) & 103.8 & 118.5\\
\bottomrule
\end{tabular}}\color{black}
\end{table}

\subsection{Scenarios and architecture}
To achieve meaningful results, it is necessary to define the scenarios and the PQC schemes to be evaluated.
In order to analyze different conditions, two distinct scenarios are considered: an urban environment (characterized by high vehicle density) and a rural environment
(with low vehicle density), both of which are implemented and available online.
This approach aims to assess how PQC algorithms perform under varying levels of network stress.

The Veins-Inet example consists of three vehicles. One of them is involved in an accident and subsequently broadcasts a warning message to the others.
Upon receiving this message, the remaining vehicles adjust their trajectories accordingly.
As for the Veins example scenario, {65 vehicles are considered during the reported experimental window, along with a single Road Side Unit (RSU).\color{black}} Following the same logic as in the V2V scenario,
when a vehicle is involved in an accident, it broadcasts an alert message to nearby vehicles and the RSU. Upon receiving
the message, the RSU relays it over a wider range, allowing vehicles that were outside the initial broadcast coverage to also receive the alert.

In the rural scenario, for Kyber, the entire encapsulation and decapsulation process was carried out, and finally, to verify its functionality, accident messages were encrypted using AES-256 with the shared secret key obtained through the process. However, in the V2X scenario, this encryption was not performed because this study aims solely to evaluate the execution times of post-quantum encryption operations, therefore, the accident messages were not encrypted. 

As for the digital signature schemes, all sent packets were signed. In both scenarios, we have two different types of packets to demonstrate their functionality: signed and unsigned. The scenario begins by sending an unsigned packet, after it is rejected by all network entities, the car that causes the accident sends its signed packet. The other vehicles receive and successfully validate the packet, and consequently change the lane in which they were traveling.

\begin{table}[H]
\centering
\caption{Parameters used for each scenario}
\begin{tabular}{|l|c|c|}
\hline

\textbf{Parameters} & \textbf{V2V} & \textbf{V2X} \\
\hline
Transmitted power & 20mW & 20mW  \\
Bandwidth        & 10MHz  & 10MHz  \\
Number of channels  & 3  &  -  \\
Standard type  & "p", IEEE 802.11p  & "p", IEEE 802.11p\\
\hline
Update interval & 0.1s  & 1s   \\
Select module type  & VeinsInetCar & TraCIDemo11p  \\
\hline
Acceleration  & 2.6 & 2.6   \\
Deceleration  &  4.5 & 4.5   \\
Sigma  & 0.5 &  0.5  \\
Maximum speed  &  14 &  14  \\
Number of vehicles  & 3  &  {65\color{black}} \\\
Starting number of vehicles  & 0 & 0 \\
Number of accidents  & 1  & 1  \\
Accident timestamp  & 20s  & 73s   \\
Accident duration  & 3s  & 20s   \\
Simulation time  & 60s  & 200s   \\
\hline
Application layer type  &  TraCIDemo11p &    \\
Header length  & 80 bits & 80 bits   \\
\hline
Maximum interference distance  & -  & 2600m  \\
Bitrate  & -  & 6Mbps  \\
Minimum power & - & -110dBm   \\
Noise floor  & - & -98dBm  \\
\hline
\end{tabular}

\begin{minipage}{9cm}
\vspace{1mm}
    {The reported urban experiments consider 65 effective vehicles. The road topology can support up to 195 vehicles during a longer simulation period.\color{black}}
  \end{minipage}

\end{table}
\begin{figure}[H]
\centering
{\subfloat[Urban environment (V2X) with RSU support]{\includegraphics[width=0.8\textwidth]{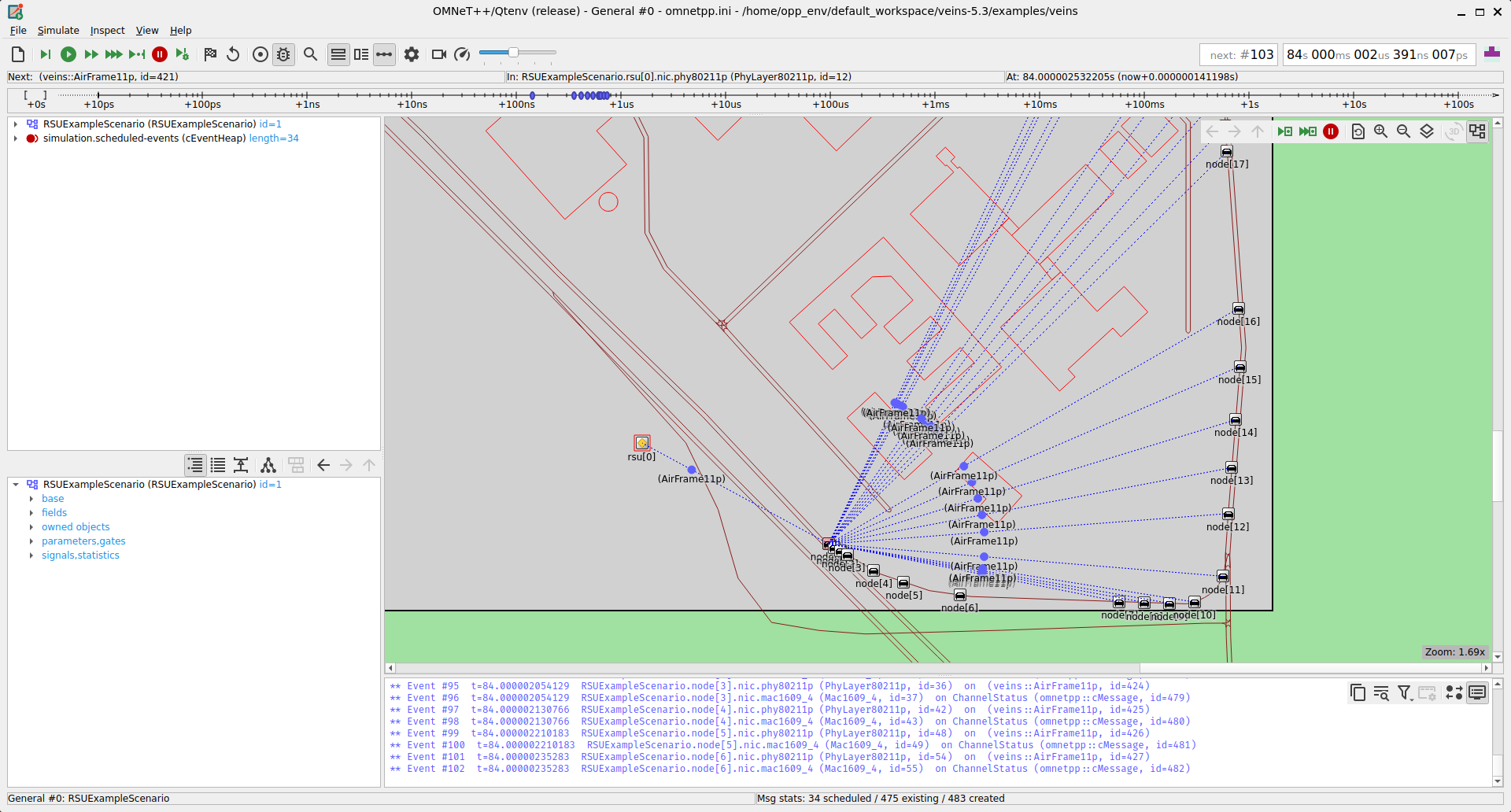}\label{fig:scenarios_urban}}\\[2mm]
\subfloat[Rural environment (V2V) with low vehicle density]{\includegraphics[width=0.98\textwidth]{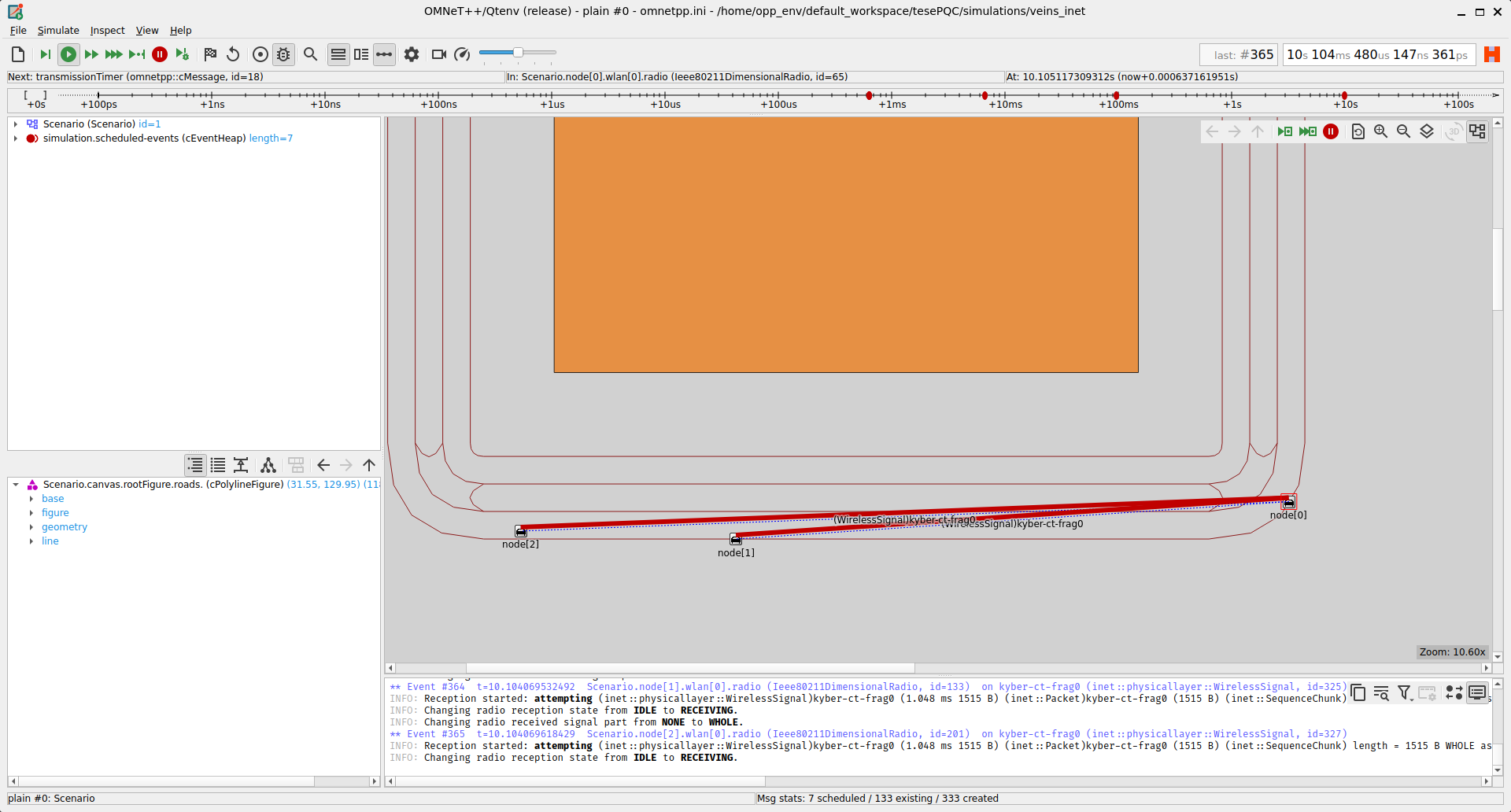}\label{fig:scenarios_rural}}}\color{black}
\caption{Developed simulation scenarios: (a) urban environment (V2X) with RSU support; (b) rural environment (V2V) with low vehicle density.}
\label{fig:scenarios}
\end{figure}

{Fig.~\ref{fig:scenarios} shows screenshots of the two simulation scenarios developed in this work, captured from the OMNeT++/Qtenv graphical environment during execution. Fig.~\ref{fig:scenarios_urban} depicts the urban V2X scenario (Veins RSU example): the road network and buildings are drawn as polygons, vehicles appear as numbered nodes (node[3] to node[17] in the captured instant), and the RSU (rsu[0]) is shown on the left; the dashed blue lines represent IEEE 802.11p frames (AirFrame11p) being broadcast simultaneously by many vehicles, which illustrates the synchronized transmission burst that causes the contention discussed in Section~IV. Fig.~\ref{fig:scenarios_rural} depicts the rural V2V scenario (Veins--INET example) with only three vehicles (node[0], node[1], and node[2]) on a single road and no infrastructure; the red lines show a fragmented Kyber ciphertext (kyber-ct-frag0) being transmitted directly between vehicles. The contrast between the two panels highlights the difference in vehicle density and network topology that drives the different channel loads reported in Table~\ref{tab:channel_metrics}.}\color{black}

\begin{figure}[H]
\centering
\includegraphics[width=5in]{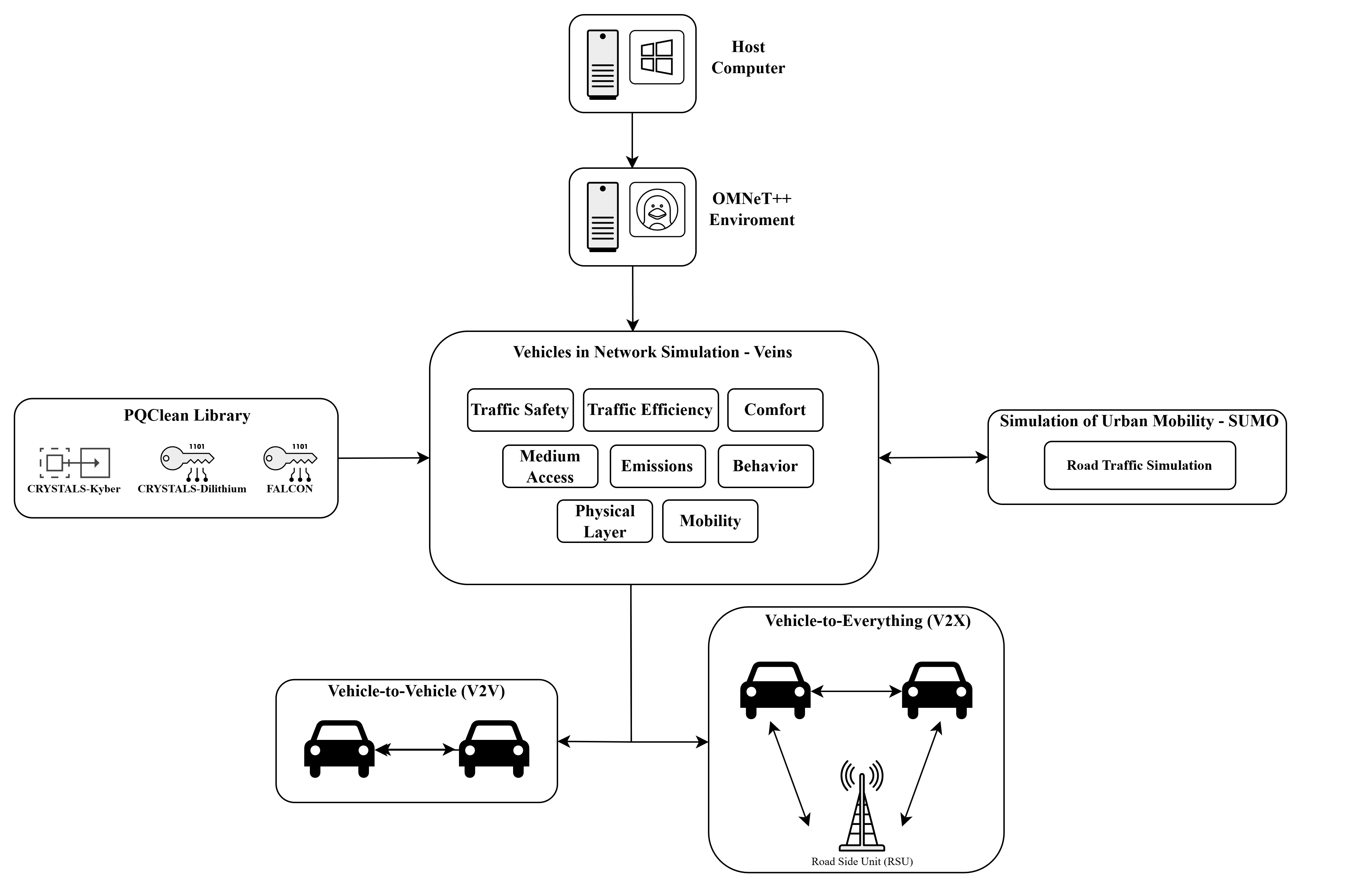}
\caption{Architecture of the experimental simulation}
\label{diagram}
\end{figure}

Figure \ref{diagram} provides a clearer overview of the architecture of this work, summarizing all the details described in this section.

\subsection{Measurement perspective and end-to-end latency decomposition}
To interpret results, it is useful to separate cryptographic computation from network effects. For a generic exchange, we can write an approximate decomposition:
\begin{equation}
T_{\text{e2e}} \approx T_{\text{crypto}} + T_{\text{MAC}} + T_{\text{prop}} + T_{\text{queue/backoff}},
\end{equation}
where $T_{\text{crypto}}$ aggregates the application-layer primitive timings (KeyGen/Encap/Decap or KeyGen/Sign/Verify), and the remaining terms capture contention, propagation, and medium access delays.
In high-density scenarios, $T_{\text{queue/backoff}}$ and collision-induced retransmissions/loss can dominate the probability of completing a handshake within a limited simulation time window.

The Rural Scenario uses Packet (VEINS+INET), while the Urban Scenario uses TraCIDemo11pMessage (VEINS), which inherits from BaseFrame1609-4. 
INET Packet: The container for the payload and the central header, designed to be transmitted across virtually all layers of the OSI model (from the application layer to the physical layer, including the network layer).
TraCIDemo11pMessage: A message strictly at the application layer (representing a short WAVE message). It is encapsulated in VEINS MAC-layer frames, such as BaseFrame1609-4 and AirFrame11p, for wireless transmission.

We observed that, in the rural scenario, the heavier schemes such as Kyber-1024, Dilithium-88, and FALCON-1024 needed to fragment their packets, since the packet size exceeded the transmission capacity. In contrast, the fragmentation process in Airframe is performed at the MAC layer before reaching the physical layer. 
The packet sizes (accident information size, signature size, public key size) for Dilithium-44, 65, and 87 are 3740, 5269, and 7227, respectively. Those for Falcon-512 and 1024 are 1561 and 3074. 
\color{black}

\section{Results}
{This section presents and analyzes the experimental results obtained from ML-KEM/CRYSTALS-Kyber, ML-DSA/CRYSTALS-Dilithium, and the NIST-selected FALCON candidate across two vehicular communication scenarios: a rural V2V environment and an urban V2X environment.\color{black}} All measurements were collected on two distinct hardware platforms (described in Section~III), allowing a cross-device comparison. The results are organized by scheme type: first the KEM scheme (Kyber), then the digital signature schemes (Dilithium and Falcon). For each scheme, we analyze primitive-level timings (in microseconds), scenario-level behavior, and the interplay between cryptographic cost and network dynamics. The goal is to assess not only raw computational performance but also the practical feasibility of each scheme in safety-critical vehicular communication. We also evaluated metrics such as median, standard deviation and confidence intervals (95\%), all reported in \url{https://github.com/MarianaCGoncalves/V2X-PQC-Tese-Results}. 

\subsection{Overview of measured operations}
We report microsecond-level timings for primitive operations (KeyGen, Encap, Decap, Sign, Verify) and relate them to scenario-level behavior.
In practice, even if cryptographic operations are fast, the \emph{communication pattern} (broadcast vs point-to-point, number of participants, synchronized transmissions)
can substantially affect the success rate of key establishment and message delivery.
It is important to note that the timings reported here reflect the execution of cryptographic primitives within the simulation environment, running on top of the host operating system via WSL2.
As such, they are influenced by OS scheduling, memory access patterns, and CPU micro architecture, rather than representing bare-metal embedded performance.
Nevertheless, they provide a meaningful relative comparison between schemes and hardware platforms, and serve as an upper-bound estimate for the cryptographic overhead in a realistic vehicular deployment.

\subsection{Kyber (KEM) results: rural (V2V) vs urban (V2X)}
In V2V, every vehicle acts like trust center, they send each other the public key and ciphertext.
The vehicle which sends the accident alert, encrypts the message with AES-256 with the now shared secret. The next tables present Hardware 1 (above) and 2 (under) results.

\begin{table}[H]
\caption{Kyber Rural Scenario}
\centering
\label{kyber_v2v}
\footnotesize
\begin{tabular}{lccc}
\toprule
Scheme & Key Gen ($\mu$s) & Encap ($\mu$s) & Decap ($\mu$s) \\
\midrule
Kyber-512  & 46.33 & 53.28 & 62.33 \\
Kyber-768  & 76.56 & 75.89 & 92.11 \\
Kyber-1024 & 106.56 & 119.11 & 133.33 \\
\midrule
Kyber-512  & 39.66 & 49 & 63.22 \\
Kyber-768  & 61.11 & 70.44 & 88.72 \\
Kyber-1024 & 99 & 106.11 & 130.72 \\
\bottomrule
\end{tabular}
\end{table}

The rural scenario exhibits the expected scaling with the security level: higher parameter sets (e.g., Kyber-1024) increase KeyGen/Encap/Decap time.
Nevertheless, these operations remain at the order of tens to hundreds of microseconds, suggesting that the cryptographic computation itself is not the limiting factor for many V2V settings.
The difference between hardware platforms is consistent with CPU capability: the second hardware yields lower or comparable timings in most operations.
Notably, Kyber-512 on Hardware 2 achieves a KeyGen time of only 39.66~$\mu$s and an Encap of 49~$\mu$s, which are well within the timing budgets of safety-critical V2V applications.
Even at the highest security level (Kyber-1024), the total KEM cost (KeyGen + Encap + Decap) remains below 340~$\mu$s on Hardware 1 and below 336~$\mu$s on Hardware 2, confirming that Kyber introduces negligible cryptographic overhead relative to network-level delays.

In V2X, the RSU acts like a trust center by generating a key pair. It then broadcasts its public key,
then vehicles proceed with the encapsulation of ciphertext with the received public key.
And concluding this KEM system, the RSU receives the ciphertext and would decapsulate, and both parties would have access to the secret.
This one-to-many key establishment pattern is fundamentally different from V2V: a single RSU public key broadcast triggers a simultaneous encapsulation response from all vehicles in range, creating a synchronized transmission burst.

As the number of vehicles in the network increases, the volume of transmitted packets also grows significantly. This leads to increased packet collisions at the MAC layer (e.g., under IEEE 802.11p), resulting in the loss of multiple ciphertext packets.
This would happen because vehicles would send them simultaneously after receiving the broadcast public key. It created bursts of transmissions and would flood the channel and because
of that many handshakes wouldn't be completed.
To mitigate this issue, it was necessary to induce a random transmission interval before sending the ciphertexts, that way it lowered the probability of collisions.
Before introducing the random interval, the handshake success rate was 23.07\%, and with the 40\% increase, the success rate rose to 63.07\%. Many of the vehicles were unable to complete their handshake due to the simulation time.
Initially, only 15 of the 65 vehicles were able to complete the handshake, and after introducing the random interval, the number of vehicles increased to 41.

The latency considers the network, MAC, propagation and random backoff of the scenario, overall latency measured 11.04~ s (Hardware 1) and 9.87~ s (Hardware 2).
This end-to-end latency is dominated by MAC-layer dynamics rather than cryptographic computation, reinforcing that Kyber's primitive cost is not the bottleneck in urban V2X deployments.

\begin{table}[H]
\caption{Kyber Urban Scenario}
\centering
\label{kyber_v2x}
\footnotesize
\begin{tabular}{lccc}
\toprule
Scheme & Key Gen ($\mu$s) & Encap ($\mu$s) & Decap ($\mu$s)  \\
\midrule
Kyber-512  & 49.67 & 63.98 & 68.81 \\
Kyber-768  & 72.00 & 87.54 & 109.41 \\
Kyber-1024 & 111.33 & 133.93 & 153.33\\
\midrule
Kyber-512  & 47.66 & 55.27 & 62.30 \\
Kyber-768  & 69.66 & 77.56 & 101.28 \\
Kyber-1024 & 90.66 & 118.97 & 135.40\\
\bottomrule
\end{tabular}
\end{table}

Comparing Tables~\ref{kyber_v2v} and~\ref{kyber_v2x}, the urban scenario shows slightly higher primitive timings than the rural one, particularly for Encap and Decap.
This is consistent with increased system load: the simulation host is simultaneously managing a larger number of vehicles, RSU logic, and network events, which introduces additional scheduling jitter.
The increase is modest (typically 5--15\% across parameter sets), confirming that Kyber's computational cost scales gracefully with scenario complexity.
Although Kyber computation remains microsecond-level, the urban scenario demonstrates that \emph{system-level feasibility} also depends on how many devices attempt to complete the exchange concurrently.
In a broadcast-triggered encapsulation wave, many vehicles may respond within the same MAC contention window, amplifying collisions and losses. This observation motivates future protocol design choices such as:
(i) randomized response timing (as done here), (ii) scheduling policies, (iii) RSU-assisted coordination, or (iv) group-key strategies, depending on application and standards constraints.

\subsection{Digital signatures: Dilithium vs Falcon under V2V/V2X}
In the digital signature scenario, once the vehicle has an accident, it signs the message and broadcasts it to all vehicles and RSU present. Once the
vehicles receive the message, they verified it, if it's valid they change the lane to avoid the accident, on the other hand, they simply drop the message.
As for the RSU, it proceeds with the same verification process, but if message is valid, it rebroadcasts it, to reach other vehicles that are further away.

From a workload perspective, this pattern emphasizes verification: a single signature may be verified by many receivers, and the verification delay can affect how quickly vehicles can react.
Therefore, while signing and key generation are important, verification cost is often the most critical metric for broadcast safety messaging.
In a scenario with $N$ vehicles and one RSU, a single signed alert message triggers up to $N$ verification operations, making the aggregate verification cost proportional to the network size.
This asymmetry between signing (one operation) and verification (many operations) is a key design consideration when selecting a signature scheme for V2X.

\begin{table}[H]
\caption{Dilithium - Rural Scenario}
\centering
\label{dilithium_v2v}
\footnotesize
\begin{tabular}{lccc}
\toprule
Scheme & Key Gen ($\mu s$) & Sign ($\mu s$) & Verify ($\mu s$) \\
\midrule
Dilithium-44 & 122,33 & 306,33 & 124,90 \\
Dilithium-65 & 221,56 & 367,67 & 200,61 \\
Dilithium-87 & 316,56 & 554,33 & 318,85 \\
\midrule
Dilithium-44 & 122,22 & 274,33 & 111,74 \\
Dilithium-65 & 167,11 & 284,67 & 178,22 \\
Dilithium-87 & 250,11 & 498,67 & 275,80 \\
\bottomrule
\end{tabular}
\end{table}

\begin{table}[H]
\centering
\caption{Falcon - Rural Scenario}
\label{falcon_v2v}
\footnotesize
\begin{tabular}{lccc}
\toprule
Scheme & Key Gen ($\mu s$) & Sign ($\mu s$) & Verify ($\mu s$) \\
\midrule
Falcon-512 & 17533 & 7186,33 & 57,76 \\
Falcon-1024 & 41544,44 & 13139 & 110,47 \\
\midrule
Falcon-512 & 14344,11 & 3848,33  & 48,28 \\
Falcon-1024 & 36992,44 & 8308,33 & 94,80 \\
\bottomrule
\end{tabular}
\end{table}

Analyzing the rural scenario results for both schemes, a clear contrast emerges. Dilithium presents well-balanced timings across all three operations: KeyGen, Sign, and Verify all remain in the hundreds of microseconds range, with Dilithium-44 being the most efficient variant. Falcon, on the other hand, exhibits a strongly asymmetric profile: KeyGen and Sign are orders of magnitude slower than Dilithium (reaching tens of milliseconds), while Verify is dramatically faster, Falcon-512 achieves verification in only 57.76~$\mu$s on Hardware 1 and 48.28~$\mu$s on Hardware 2, compared to 124.90~$\mu$s and 111.74~$\mu$s for Dilithium-44 respectively. This represents a verification speedup of approximately $2\times$ in favor of Falcon.

The latency measured for Dilithium was 2.12~s, slightly slower than FALCON. FALCON demonstrates a relatively low overall latency of 2.08~s when accounting for signing, verification, MAC operations, and networking overhead.
However, Dilithium is capable of performing more efficiently on the second hardware and outperforms FALCON with a latency of 2.0062~s, whereas FALCON has a latency of 2.0127~s.
This inversion between hardware platforms is noteworthy: it suggests that Falcon's Gaussian sampling-based signing is more sensitive to CPU micro architecture and memory bandwidth than Dilithium's rejection-sampling approach, making hardware selection a relevant factor when deploying Falcon in embedded automotive systems.

A key trade-off becomes evident: Falcon has much higher KeyGen and Sign times, but achieves faster verification than Dilithium.
In V2X broadcast settings, this can be advantageous because verification is executed far more frequently than signing. Therefore, protocol designers may accept slower signing at the sender side
if verification throughput at receivers improves and reduces message processing backlog.

The entire Kyber process takes place before the accident occurs, that is, the network is free to carry out the entire encapsulation and decapsulation process. Therefore, we begin measuring latency when the vehicle receives the RSU’s public key and begins encapsulation, and it ends at the RSU when it completes decapsulation.
This is in contrast to digital signature schemes, where latency measurement begins when the accident message is signed by the vehicle and ends when the RSU validates the message. 

\begin{figure}[H]
\centering
\footnotesize
\includegraphics[width=2.6in]{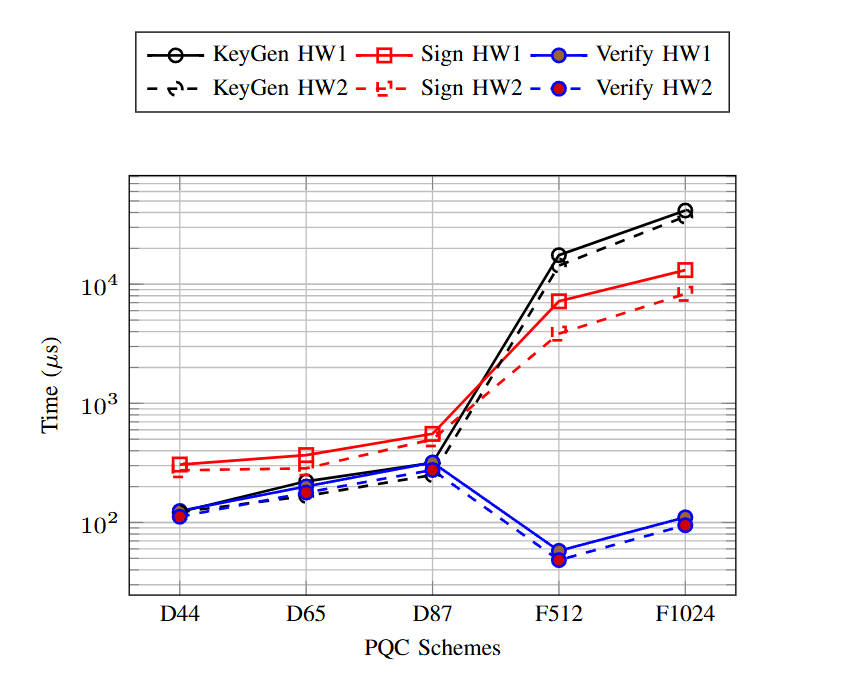}
\caption{Comparison of both hardware on Rural scenario (V2V)}
\label{fig:performance_v2v}
\end{figure}

When it comes to the automotive context, we must ensure that the schemes do not disturb the fundamental functionalities within a vehicle.
As observed in the simulated scenario, every cryptographic function operates in the order of microseconds (or a few milliseconds for Falcon signing/keygen), which is a good sign for V2X feasibility at the primitive level.
Kyber could be a good candidate to cooperate with Dilithium or FALCON, depending on whether the target application prioritizes simpler implementations and balanced costs (Dilithium)
or faster verification in broadcast-heavy workloads (Falcon).

\begin{table}[H]
\centering
\caption{Dilithium - Urban Scenario}
\label{dilithium_v2x}
\footnotesize
\begin{tabular}{lccc}
\toprule
Scheme & Key Gen ($\mu s$) & Sign ($\mu s$) & Verify ($\mu s$)\\
\midrule
Dilithium-44 & 122,79 & 574,17 & 137,38 \\
Dilithium-65 & 206,95 & 796,67 & 198,91 \\
Dilithium-87 & 343,84 & 1394 & 345,39 \\
\midrule
Dilithium-44 & 108,70 & 345,83 & 140,20 \\
Dilithium-65 & 180,67 & 768,20 & 193,99 \\
Dilithium-87 & 277,30 & 1450 & 267,16 \\
\bottomrule
\end{tabular}
\end{table}

\begin{table}[H]
\centering
\caption{Falcon - Urban Scenario}
\label{falcon_v2x}
\footnotesize
\begin{tabular}{lccc}
\toprule
Scheme & Key Gen ($\mu s$) & Sign ($\mu s$) & Verify ($\mu s$)\\
\midrule
Falcon-512 & 17316,15 & 4936,67 & 69,49 \\
Falcon-1024 & 86291,81 & 10849,83 & 110,55 \\
\midrule
Falcon-512 & 12083,80 & 4086,5 & 64,10 \\
Falcon-1024 & 38437,77 & 8646,5 & 118,50 \\
\bottomrule
\end{tabular}
\end{table}

{The urban scenario results reveal an important effect of increased network load on the wall-clock timings observed during simulation execution. In particular, Falcon-1024 KeyGen rises substantially from the rural to the urban scenario on Hardware 1, despite key generation being algorithmically scenario-independent. This difference is attributed to host-load contamination: OMNeT++, Veins, INET, SUMO, vehicle modules, RSU logic, and network events execute on a shared WSL2 host, where CPU scheduling, cache contention, and operating-system interference can affect wall-clock measurements. Cross-scenario primitive-level comparisons must therefore be interpreted cautiously.\color{black}}

The urban scenario results reveal an important effect of increased network load on signing performance. Comparing Dilithium's Sign times between rural (Table~\ref{dilithium_v2v}) and urban (Table~\ref{dilithium_v2x}) scenarios, a significant increase is observed: Dilithium-44 Sign rises from 306.33~$\mu$s to 574.17~$\mu$s on Hardware 1, and Dilithium-87 Sign reaches 1394~$\mu$s in the urban case. This degradation is attributable to the higher simulation load in the urban environment, where more concurrent events compete for CPU time. Falcon's signing also increases in the urban scenario, though its verification times remain consistently low and stable across both scenarios, further reinforcing its suitability for broadcast-heavy V2X workloads.

But even though FALCON takes more time in generating keys and signing, its verification average always beats Dilithium.
So creating a protocol with Kyber and FALCON would actually work very efficiently. Since the verification function is the more frequent operation during V2V and V2X communications,
it would be reasonable to make the effort with FALCON: it only generates keys once, and although signing is more time-consuming, the verification time could be very beneficial especially in V2X scenarios.
Comparing the two scenarios, V2X requires slightly more time than V2V because V2X has more identities interacting with each other, affecting the latency.
In addition, higher density increases the probability that cryptographic processing and MAC contention overlap, creating queues and potentially delaying safety reactions if not engineered carefully.

\begin{figure}[H]
\centering
\footnotesize
\includegraphics[width=2.5in]{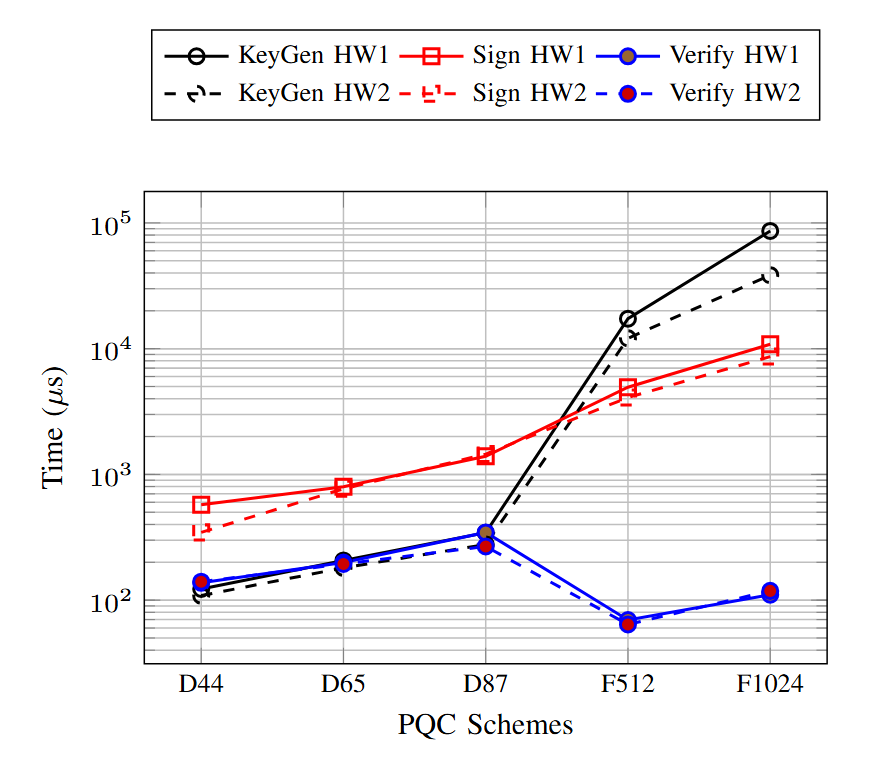}
\caption{Comparison of both hardwares on Urban scenario (V2X)}
\label{fig:performance_v2x}
\end{figure}

As we observe in Figures~\ref{fig:performance_v2v} and~\ref{fig:performance_v2x}, Hardware 2 (Table~\ref{tab:hardware1}) delivers better performance in virtually all operations. This is consistent with the fact that the AMD Ryzen 5 5500U processor in Hardware 2 offers a more modern microarchitecture and higher memory bandwidth than the Intel i5-8400 in Hardware 1, resulting in lower execution times particularly for memory-intensive operations such as Falcon's key generation and Dilithium's signing. These hardware-dependent differences highlight the importance of evaluating PQC schemes across multiple platforms, as performance rankings between schemes can shift depending on the underlying processor.

\section{Discussion and conclusions}
This paper evaluated the performance of lattice-based PQC schemes: CRYSTALS-Kyber, CRYSTALS-Dilithium, and FALCON in V2X and V2V communication scenarios as developed in this paper.
The analysis considered both rural (V2V) and urban (V2X) environments, with the aim of assessing performance under varying network conditions and levels of communication stress.
The results indicate that Kyber is a strong candidate for integration into future protocols, due to its consistently favorable performance across both scenarios.
Regarding digital signature schemes, Dilithium and FALCON exhibit distinct trade-offs, however both demonstrate characteristics that make them suitable for deployment in
automotive communication systems. In particular, while FALCON achieves lower verification latency, Dilithium provides competitive performance with different design advantages.
To mitigate latency and communication overhead, periodic packet transmission is identified as a potential optimization strategy, reducing channel contention among vehicles.
The impact of PQC is more pronounced in V2X scenarios, largely due to the relative efficiency of key encapsulation mechanisms such as Kyber when
compared to signature-based approaches.
{Overall, the evaluated schemes demonstrate comparative performance characteristics that can inform future V2V and V2X protocol design. The present findings do not provide an absolute PQC-versus-classical feasibility claim because no ECDSA/ECDH baseline was measured in the same OMNeT++/Veins/INET/WSL2 environment.\color{black}}

{Additional limitations should be considered. First, primitive timings were collected as wall-clock measurements while the simulation stack executed on a shared host; therefore, scenario-to-scenario differences in primitive timings may include host-load contamination rather than intrinsic cryptographic variation. Second, the study does not model IEEE 1609.2 or ETSI TS 103 097 certificate chains, certificate digests, or pseudonym management, which are relevant contributors to practical V2X communication overhead. Third, CBR, standalone PDR, and MAC-layer collision probability require extraction from the final simulation logs. Finally, future work must include dedicated-core cryptographic benchmarks and ECDSA/ECDH measurements in the same simulation setup to establish a controlled conventional baseline.\color{black}}

The main limitation of this paper is the use of a virtual simulation environment. Although the Veins framework enables the modeling of
complex scenarios, it cannot fully capture real-world problems such as signal interference, unpredictable channel conditions, and hardware-specific constraints.
Consequently, the results may not fully reflect performance in real deployment settings.
A second limitation concerns the simplicity of the V2X scenario considered. The simulated environment includes only vehicles and a roadside unit (RSU), restricting communication
patterns to V2V and V2I interactions. This abstraction does not account for more heterogeneous and large-scale V2X ecosystems, which may include additional infrastructure elements,
pedestrians and network.

This work only tested with lattice-based schemes due to their popularity among PQC schemes, because of their mathematical complexity can garantee security.
But that leaves us with still multiple PQC schemes, like SPHINCS$^+$ (Hash-based), HQC (Code-based) which is KEM whose
security is based on the hardness of solving the Quasi-Cyclic Syndrome Decoding (QCSD) problem, Saber (Lattice-based) also a KEM that uses Learning with Rounding (LWR).
Now is the time to start and experimenting like the authors of \cite{ref45} which joined two schemes: Kyber and Saber to create a efficient protocol. Not only
joining two PQC schemes but also creating hybrid protocols which can operate at all situations, following the author's methodology in \cite{v2x_pqc_in_protocols} in which they explain each step of implementing these PQC protocols in V2X environment.

With the raise of Artificial Intelligence it is only natural to implement it in this context. Complementing the idea of analyzing and testing with every PQC scheme,
AI-based adaptive security could monitor performance indicators and select cryptographic policies based on system conditions and compliance requirements \cite{NOETZOLD2026131705,11488882,NOETZOLD2026131705}.
Related directions include end-to-end simulation and predictive analysis frameworks for IoT/smart environments \cite{11488882}, blockchain-based authentication with post-quantum security \cite{latif2026},
and AI-powered security classification in post-quantum lightweight settings \cite{mumtaz2026}. Although these works do not directly target V2X, they support a broader trend toward
adaptive, data-driven security mechanisms that can be relevant for future connected mobility systems.

The study \cite{ref30} in Related Work section only tests the S32G. The hardware mentioned is PQC-recommended by NPX, so it would be very compelling to analyze the results.
On top of everything, we still have to deal with attacks of every kind, as reported in papers like \cite{ref36} and \cite{scas_conclusion} which alert about the efficiency
of side-channel attacks on schemes like Kyber, Saber, Dilithium, HQC, we need to be cautions by applying the appropriate countermeasures.

Finally, this work produced a dataset \cite{ref50} comprising key generation, encapsulation/decapsulation and signing/verification timings for each scheme. This dataset provides a foundation
for future research, including the application of machine learning techniques to dynamically select the most appropriate cryptographic scheme based on system conditions. Other scenarios and techniques may also be included, as presented in the studies \cite{Valderi}, \cite{Sergio}, \cite{2026} and \cite{Dinis}.

\section*{Acknowledgments}
This research was partially funded by national funds through the FCT - Foundation for Science and Technology, I.P. within the scope of projects UIDB/04466/2025 and UIDP/04466/2025. This research was also partially funded by national funds through the FCT - Foundation for Science and Technology, I.P. within the scope of project 16881, LISBOA2030-FEDER-00816400, with DOI: https://doi.org/10.54499/2023.16583.ICDT.

\bibliographystyle{plainnat}
\bibliography{references}

@online{ref1,
  author    = {Swayne, M.},
  year      = {2025},
  title     = {Quantum Computing Roadmaps \& Predictions of leading players},
  note       = {[Online]: \url{https://thequantuminsider.com/2025/05/16/quantum-computing-roadmaps-a-look-at-the-maps-and-predictions-of-major-quantum-players/}},
  publisher = {Quantum Insider}
}

@online{ref2,
  author    = {Zoting, S.},
  year      = {2025},
  title     = {Automotive PQC Market Driven by Rising Cybersecurity Threats and Quantum-Safe Certification},
  note       = {[Online]: https://www.precedenceresearch.com/automotive-pqc-market},
  publisher = {Precedence Research}
}

@online{ref3,
  author    = {Danish Quantum Community},
  year      = {2025},
  title     = {Denmark to acquire the World´s Most Powerful Quantum Computer},
  note       = {[Online]: \url{https://dqc.dk/denmark-to-acquire-the-worlds-most-powerful-quantum-computer/}},
  publisher = {Danish Quantum Community}
}

@misc{ref4,
	title = {Elliptic {Curve} {Cryptography} in {Vehicle} {Security}},
	copyright = {https://creativecommons.org/licenses/by/4.0/},
	url = {https://www.techrxiv.org/users/876135/articles/1255855-elliptic-curve-cryptography-in-vehicle-security?commit=e6288063439aeed5c7162fdef5e469ffa20662c7},
	doi = {10.36227/techrxiv.173611692.29943420/v1},
	urldate = {2026-01-27},
	publisher = {Preprints},
	author = {Manson, Erin and Mood, Ryan},
	month = jan,
	year = {2025},
}

@online{ref6,
  author    = {NIST},
  year      = {2022},
  title     = {Post-Quantum Cryptography},
  note      = {[Online]: \url{https://csrc.nist.gov/projects/post-quantum-cryptography/post-quantum-cryptography-standardization/selected-algorithms}},
  publisher = {NIST}
}

@online{ref27,
  author    = {Aaron Rathweg and Abdelrahman Osman and Andreas Fleig and Efstathia Katsigianni and Alexander Tschache},
  year      = {2024},
  title     = {Impacts of Post-Quantum Cryptography on Automotive Security: A Case Study},
  doi = {10.13154/294-10394},
  url       = {https://d-nb.info/1316435830/34},
  note      = {Accessed: 12 November 2025},
  publisher = {Deutsche Nationalbibliothek}
}

@misc{ref28,
  author = {Wen Wang and Marc Stöttinger},
  title = {Post-Quantum Secure Architectures for Automotive Hardware Secure Modules},
  howpublished = {Cryptology ePrint Archive, Paper 2020/026},
  year = {2020},
  note = {Available in: \url{https://eprint.iacr.org/2020/026}. Accessed: June 2026}
}

@misc{ref30,
  author = {Joppe W. Bos and Alexander Dima and Alexander Kiening and Joost Renes},
  title = {Post-Quantum Secure Over-the-Air Update of Automotive Systems},
  doi = {10.13154/294-10380},
  howpublished = {Cryptology {ePrint} Archive, Paper 2023/965},
  year = {2023},
  url = {https://eprint.iacr.org/2023/965}
}

@inproceedings{ref31,
  address = {San Diego, CA, USA},
	title = {When {Cryptography} {Needs} a {Hand}: {Practical} {Post}-{Quantum} {Authentication} for {V2V} {Communications}},
	isbn = {978-1-891562-93-8},
	shorttitle = {When {Cryptography} {Needs} a {Hand}},
	url = {https://www.ndss-symposium.org/wp-content/uploads/2024-267-paper.pdf},
	doi = {10.14722/ndss.2024.24267},
	language = {en},
	urldate = {2026-01-11},
	booktitle = {Proceedings 2024 {Network} and {Distributed} {System} {Security} {Symposium}},
	publisher = {Internet Society},
	author = {Twardokus, Geoff and Bindel, Nina and Rahbari, Hanif and McCarthy, Sarah},
	year = {2024},
}

@article{ref32,
  title = {A {Review} on {Security} {Challenges} in {V2X} {Communications} {Technology} for {VANETs}},
	volume = {13},
	copyright = {https://creativecommons.org/licenses/by/4.0/legalcode},
	issn = {2169-3536},
	url = {https://ieeexplore.ieee.org/document/10879476/},
	doi = {10.1109/ACCESS.2025.3541035},
	urldate = {2026-01-11},
	journal = {IEEE Access},
	author = {Farsimadan, Eslam and Moradi, Leila and Palmieri, Francesco},
	year = {2025},
	pages = {31069--31094},
}

@misc{ref33,
  title = {Evaluation of {Jamming} {Attacks} on {NR}-{V2X} {Systems}: {Simulation} and {Experimental} {Perspectives}},
	copyright = {http://creativecommons.org/licenses/by/4.0},
	shorttitle = {Evaluation of {Jamming} {Attacks} on {NR}-{V2X} {Systems}},
	url = {https://www.preprints.org/manuscript/202507.0729/v1},
	doi = {10.20944/preprints202507.0729.v1},
	urldate = {2026-01-08},
	publisher = {Engineering},
	author = {Da Silva, Antonio Santos and Gularte, Kevin Herman Muraro and Santos, Giovanni Almeida and Corrêa, Davi Salomão Soares and Oliveira De Melo, Luís Felipe and Da Costa, João Paulo Javidi and Vargas, José Alfredo Ruiz and Da Silva, Daniel Alves and Fei, Tai},
	month = jul,
	year = {2025},
}

@misc{ref36,
  author = {Abiodun Olaluwe and Nouf Nur Nabilah and Sheikh Tareq and Akshay Raghavendra Kulkarni and Annamalai Annamalai},
  title = {Machine Learning and Side-Channel Attacks on Post-Quantum Cryptography},
  howpublished = {Cryptology ePrint Archive, Paper 2025/1754},
  year = {2025},
  note = {Available in: \url{https://eprint.iacr.org/2025/1754}. Accessed: June 2026}
}

@inproceedings{ref37,
  address = {Kranj, Slovenia},
	title = {Strengthening {Post}-{Quantum} {Security} for {Automotive} {Systems}},
	copyright = {https://ieeexplore.ieee.org/Xplorehelp/downloads/license-information/IEEE.html},
	isbn = {978-1-7281-9535-3},
	url = {https://ieeexplore.ieee.org/document/9217638/},
	doi = {10.1109/DSD51259.2020.00094},
	urldate = {2026-01-11},
	booktitle = {2020 23rd {Euromicro} {Conference} on {Digital} {System} {Design} ({DSD})},
	publisher = {IEEE},
	author = {Fritzmann, Tim and Vith, Jonas and Sepulveda, Johanna},
	month = aug,
	year = {2020},
	pages = {570--576},
}

@misc{ref38,
  title = {Post-{Quantum} {Cryptography} for {Intelligent} {Transportation} {Systems}: {An} {Implementation}-{Focused} {Review}},
	shorttitle = {Post-{Quantum} {Cryptography} for {Intelligent} {Transportation} {Systems}},
	url = {http://arxiv.org/abs/2601.01068},
	doi = {10.48550/arXiv.2601.01068},
	urldate = {2026-01-12},
	publisher = {arXiv},
	author = {Mamun, Abdullah Al and Abrar, Akid and Rahman, Mizanur and Salek, M. Sabbir and Chowdhury, Mashrur},
	month = jan,
	year = {2026},
	note = {arXiv:2601.01068 [cs]},
}

@online{ref39,
  author    = {Michele La Manna and Pericle Perazzo and Luigi Treccozzi and Gianluca Dini},
  year      = {2021},
  title     = {Assessing the Cost of Quantum Security for Automotive Over-The-Air Updates},
  doi = {10.1109/ISCC53001.2021.9631426},
  url       = {https://ieeexplore.ieee.org/document/9631426},
  note      = {Accessed: 14 January 2026},
  publisher = {ieeexplore}
  
}

@article{ref40,
  title = {Securing in-vehicle communications through post-quantum cryptography and network segmentation},
	volume = {126},
	issn = {00457906},
	url = {https://linkinghub.elsevier.com/retrieve/pii/S0045790625004318},
	doi = {10.1016/j.compeleceng.2025.110488},
	language = {en},
	urldate = {2026-01-12},
	journal = {Computers and Electrical Engineering},
	author = {Castiglione, Arcangelo and Elia, Teresa},
	month = aug,
	year = {2025},
	pages = {110488},
}

@article{ref41,
  title = {A {Survey} of {Security} and {Privacy} {Issues} in {V2X} {Communication} {Systems}},
	volume = {55},
	issn = {0360-0300, 1557-7341},
	url = {https://dl.acm.org/doi/10.1145/3558052},
	doi = {10.1145/3558052},
	language = {en},
	number = {9},
	urldate = {2026-01-13},
	journal = {ACM Computing Surveys},
	author = {Yoshizawa, Takahito and Singelée, Dave and Muehlberg, Jan Tobias and Delbruel, Stephane and Taherkordi, Amir and Hughes, Danny and Preneel, Bart},
	month = sep,
	year = {2023},
	pages = {1--36},
}

@misc{ref42,
  title = {{AI}-{Driven} {Post}-{Quantum} {Cryptography} for {Cyber}-{Resilient} {V2X} {Communication} in {Transportation} {Cyber}-{Physical} {Systems}},
	copyright = {Creative Commons Attribution Non Commercial No Derivatives 4.0 International},
	url = {https://arxiv.org/abs/2510.08496},
	doi = {10.48550/ARXIV.2510.08496},
	urldate = {2026-01-14},
	publisher = {arXiv},
	author = {Abrar, Akid and Dasgupta, Sagar and Rahman, Mizanur and Alsharif, Ahmad},
	year = {2025},
	note = {Version Number: 1}
}

@article{ref43,
	title = {{PQ}-{V2X}: {A} {Novel} {Postquantum} {Cryptographic} {Dataset} for {Secure} {Vehicular} {Communications}},
	volume = {12},
	copyright = {https://ieeexplore.ieee.org/Xplorehelp/downloads/license-information/IEEE.html},
	issn = {2327-4662, 2372-2541},
	shorttitle = {{PQ}-{V2X}},
	url = {https://ieeexplore.ieee.org/document/11193862/},
	doi = {10.1109/JIOT.2025.3618153},
	number = {24},
	urldate = {2026-01-28},
	journal = {IEEE Internet of Things Journal},
	author = {Abdel Hakeem, Shimaa A. and Kim, HyungWon},
	month = dec,
	year = {2025},
	pages = {53704--53729},
}

@article{ref44,
	title = {Post-quantum cryptography for automotive systems},
	volume = {87},
	issn = {01419331},
	url = {https://linkinghub.elsevier.com/retrieve/pii/S0141933121005299},
	doi = {10.1016/j.micpro.2021.104379},
	language = {en},
	urldate = {2026-01-28},
	journal = {Microprocessors and Microsystems},
	author = {Fritzmann, Tim and Vith, Jonas and Flórez, Daniel and Sepúlveda, Johanna},
	month = nov,
	year = {2021},
	pages = {104379},
}

@article{ref45,
	title = {A {Lightweight} {Key} {Agreement} {Protocol} for {V2X} {Communications} {Based} on {Kyber} and {Saber}},
	volume = {25},
	issn = {1424-8220},
	url = {https://www.mdpi.com/1424-8220/25/22/6938},
	doi = {10.3390/s25226938},
	language = {en},
	number = {22},
	urldate = {2026-01-28},
	journal = {Sensors},
	author = {Dai, Yinfei and Wang, Qi and Song, Xiao and Wang, Shaoqiang},
	month = nov,
	year = {2025},
	pages = {6938},
}

@incollection{ref46,
	address = {Cham},
	title = {Post-quantum {Cryptography} for {Connected} and {Cooperative} {Automated} {Mobility}: {A} {Comprehensive} {Overview}},
	isbn = {978-3-032-06762-3 978-3-032-06763-0},
	shorttitle = {Post-quantum {Cryptography} for {Connected} and {Cooperative} {Automated} {Mobility}},
	url = {https://link.springer.com/10.1007/978-3-032-06763-0\_105},
	doi = {10.1007/978-3-032-06763-0\_105},
	language = {en},
	urldate = {2026-01-28},
	booktitle = {Transport {Transitions}: {Advancing} {Sustainable} and {Inclusive} {Mobility}},
	publisher = {Springer Nature Switzerland},
	author = {Mohamed, Mohamed Saied and Godard, Julie and Jimenez, Victor and Jousse, Adrien and Paños, Pau Perea and Zhang, Miao},
	editor = {McNally, Ciaran and Carroll, Páraic and Martinez-Pastor, Beatriz and Ghosh, Bidisha and Efthymiou, Marina and Valantasis-Kanellos, Nikolaos},
	year = {2026},
	note = {Series Title: Lecture Notes in Mobility},
	pages = {738--744},
}

@misc{ref50,
  author    = {Mariana Gonçalves and Valderi Leithardt},
  title     = {Post-Quantum Cryptography in Automotive Scenarios (V2X and V2V)},
  year      = {2026},
  url       = {https://data.mendeley.com/datasets/czx6vk5hsr/1},
  doi       = {10.17632/czx6vk5hsr.1},
  publisher = {Mendeley Data},
}

@inproceedings{ref52,
  author    = {Matthias J. Kannwischer and
               Peter Schwabe and
               Douglas Stebila and
               Thom Wiggers},
  title     = {Improving Software Quality in Cryptography Standardization Projects},
  booktitle = {{IEEE} European Symposium on Security and Privacy, EuroS{\&}P 2022 - Workshops, Genoa, Italy, June 6-10, 2022},
  pages     = {19--30},
  publisher = {IEEE Computer Society},
  address   = {Los Alamitos, CA, USA},
  year      = {2022},
  url       = {https://eprint.iacr.org/2022/337},
  doi       = {10.1109/EuroSPW55150.2022.00010},
}

@misc{vasconcelos_quantum_2020,
	title = {Quantum {Computing} @ {MIT}: {The} {Past}, {Present}, and {Future} of the {Second} {Revolution} in {Computing}},
	shorttitle = {Quantum {Computing} @ {MIT}},
	doi = {10.48550/arXiv.2002.05559},
	publisher = {arXiv},
	author = {Vasconcelos, Francisca},
	month = feb,
	year = {2020},
	note = {arXiv:2002.05559 [physics.pop-ph]}, 
}

@article{v2x_,
author = {Bintoro, Ketut},
year = {2025},
month = {06},
pages = {31-41},
title = {Overview of Vehicle-to-Everything Communication: Technological Advancements, Applications, and Future Prospects in Intelligent Transportation Systems},
volume = {8},
journal = {JISA(Jurnal Informatika dan Sains)},
doi = {10.31326/jisa.v8i1.2187}
}

@inproceedings{v2x_pqc_in_protocols,
author = {Luo, Ge and Liu, Jianxin},
title = {Post-Quantum Cryptography Challenges in Connected Vehicles with V2X},
year = {2024},
isbn = {9798400716959},
publisher = {Association for Computing Machinery},
address = {New York, NY, USA},
url = {https://doi.org/10.1145/3673277.3673313},
doi = {10.1145/3673277.3673313},
booktitle = {Proceedings of the 2024 3rd International Conference on Cryptography, Network Security and Communication Technology},
pages = {205–208},
numpages = {4},
location = {Harbin, China},
series = {CNSCT '24}
}

@article{scas_conclusion,
author = {Delgado-Vargas, Kevin and Mancillas-López, Cuauhtemoc and Gallegos-García, Gina},
year = {2024},
month = {12},
pages = {},
title = {A Look at Side Channel Attacks on Post-quantum Cryptography},
volume = {28},
journal = {Computación y Sistemas},
doi = {10.13053/cys-28-4-5130}
}

@article{article_v2x,
author = {Špitálová, Zuzana},
year = {2023},
month = {01},
pages = {C24-C35},
title = {Vehicle-to-Everything Communication},
volume = {25},
journal = {Communications - Scientific letters of the University of Zilina},
doi = {10.26552/com.C.2023.017}
}

@misc{valderi_ref_1,
author = {Noetzold, Darlan and Leithardt, Valderi},
year = {2026},
month = {01},
pages = {},
title = {Fundamentals, Recent Advances, and Challenges Regarding Cryptographic Algorithms for the Quantum Computing Era},
doi = {10.48550/arXiv.2601.18413}
}

@misc{overview_pqc,
      author = {Seyed MohammadReza Hosseini and Hossein Pilaram},
      title = {A Comprehensive Review of Post-Quantum Cryptography: Challenges and Advances},
      howpublished = {Cryptology {ePrint} Archive, Paper 2024/1940},
      year = {2024},
      url = {https://eprint.iacr.org/2024/1940},
      doi = {10.5267/j.ijdns.2024.12.003}
}

@article{NOETZOLD2026131705,
   author  = {Noetzold, D. and Leithardt, V. R. Q. and de Paz Santana, J. F. and Barbosa, J. L. V.},
   title   = {Oraculum: A model for self-adaptive system optimization in smart environments},
   journal = {Expert Systems with Applications},
   pages   = {131705},
   year    = {2026},
   doi     = {10.1016/j.eswa.2026.131705}
 }

@article{11488882,
   author  = {Noetzold, D. and Barbosa, J. L. V. and De Paz, J. F. and Leithardt, V. R. Q.},
   title   = {Reinforcement learning-based adaptive quantum-safe cryptography for {DN25}-compliant smart environments},
   journal = {IEEE Access},
   volume  = {14},
   pages   = {62087--62104},
   year    = {2026},
   doi     = {10.1109/ACCESS.2026.3685890}
 }

@article{latif2026,
   author  = {Latif, R. and Yakubu, B. M. and Jamail, N. S. M. and others},
   title   = {BBAS: A blockchain-based authentication system for e-health with multi-factor authentication, access control, and post-quantum security},
   journal = {Scientific Reports},
   year    = {2026},
   doi     = {10.1038/s41598-026-39415-5}
 }

@article{mumtaz2026,
   author  = {Mumtaz, S. and Sanam, N. and Mumtaz, F.},
   title   = {AI-Powered security classification of generator-matrix s-box for post-quantum lightweight cryptography},
   journal = {Telecommunication Systems},
   volume  = {89},
   pages   = {20},
   year    = {2026},
   doi     = {10.1007/s11235-025-01395-8}
 }

@inproceedings{ghosh2024,
   author    = {Ghosh, U. and Njilla, L. and Das, D. and Levin, E.},
   title     = {{FLAS}: A Federated Learning Framework for Adaptive Security in Edge-Driven {UAV} Networks},
   booktitle = {ICC 2024 - IEEE International Conference on Communications},
   pages     = {1298--1303},
   year      = {2024},
   doi       = {10.1109/ICC51166.2024.10622352}
 }

@article{Sohail_Noetzold_Leithardt_2026, title={PRISEC III: Dynamic Cryptographic Adaptation for Balancing Performance and Security}, volume={17}, url={https://journals-sol.sbc.org.br/index.php/jisa/article/view/6502}, DOI={10.5753/jisa.2026.6502}, abstractNote={&amp;lt;p&amp;gt;This paper introduces PRISEC III, a dynamic cryptographic framework designed to balance security and performance in heterogeneous IoT environments. Unlike static, one-size-fits-all approaches, PRISEC III employs a role-based multi-level model that adapts cryptographic strategies according to data sensitivity, device constraints, and network conditions. The framework integrates lightweight symmetric primitives for efficiency, robust asymmetric methods for secure key exchange, and hybrid schemes that combine multiple layers of protection. To enhance long-term resilience, PRISEC III also incorporates post-quantum options, ensuring preparedness against emerging cryptographic threats. Evaluation results demonstrate that the approach achieves scalable security while maintaining feasible computational costs, making it suitable for large-scale and resource-constrained IoT deployments.&amp;lt;/p&amp;gt;}, number={1}, journal={Journal of Internet Services and Applications}, author={Sohail, Humza and Noetzold, Darlan and Leithardt, Valderi Reis Quietinho}, year={2026}, month={May}, pages={174–191} }

@ARTICLE{10838531,
  author={Lohmiller, Nils and Kaniewski, Sabrina and Menth, Michael and Heer, Tobias},
  journal={IEEE Access}, 
  title={A Survey of Post-Quantum Cryptography Migration in Vehicles}, 
  year={2025},
  volume={13},
  number={},
  pages={10160-10176},
  doi={10.1109/ACCESS.2025.3528562}}

@inproceedings{Conti_2025,
   title={PQ-CAN: A Framework for Simulating Post-Quantum Cryptography in Embedded Systems},
   url={http://dx.doi.org/10.1109/ISCC65549.2025.11326414},
   DOI={10.1109/iscc65549.2025.11326414},
   booktitle={2025 IEEE Symposium on Computers and Communications (ISCC)},
   publisher={IEEE},
   author={Conti, Mauro and Marchiori, Francesco and Matarazzo, Sebastiano and Rubin, Marco},
   year={2025},
   month=July, pages={1–6} }

@misc{demir2025performanceanalysisindustrydeployment,
      title={Performance Analysis and Industry Deployment of Post-Quantum Cryptography Algorithms}, 
      author={Elif Dicle Demir and Buse Bilgin and Mehmet Cengiz Onbasli},
      year={2025},
      eprint={2503.12952},
      archivePrefix={arXiv},
      primaryClass={cs.CR},
      url={https://arxiv.org/abs/2503.12952}, 
}

@article{article_reviewer_1,
author = {Abbasi, Maryam and Cardoso, Filipe and Vaz, Paulo and Silva, Jose and Martins, Pedro},
year = {2025},
month = {05},
pages = {32},
title = {A Practical Performance Benchmark of Post-Quantum Cryptography Across Heterogeneous Computing Environments},
volume = {9},
journal = {Cryptography},
doi = {10.3390/cryptography9020032}
}

@article{article_reviewer_2,
author = {Demir, Elif and Gorgulu, Idil and Onbasli, M.},
year = {2026},
month = {01},
pages = {56099-56111},
title = {Micro and Transport Layer Security Benchmarking of Post-Quantum Versus Classical Cryptography},
volume = {14},
journal = {IEEE Access},
doi = {10.1109/ACCESS.2026.3682760}
}

@online{nist_pqc_standards_2024,
  author    = {National Institute of Standards and Technology},
  title     = {NIST Releases First 3 Finalized Post-Quantum Encryption Standards},
  year      = {2024},
  url       = {https://www.nist.gov/news-events/news/2024/08/nist-releases-first-3-finalized-post-quantum-encryption-standards},
  note      = {[Online]. Accessed: September 2026},
  publisher = {NIST}
}

@ARTICLE{Valderi,
  author={Leithardt, Valderi and Santos, Douglas and Silva, Luis and Viel, Felipe and Zeferino, Cesar and Silva, Jorge},
  journal={IEEE Latin America Transactions}, 
  title={A Solution for Dynamic Management of User Profiles in IoT Environments}, 
  year={2020},
  volume={18},
  number={07},
  pages={1193-1199},
  doi={10.1109/TLA.2020.9099759}}

@article{Dinis,
title = {AES hardware architectures and non-standard Rijndael extensions up to 1024-bit keys: An open-source FPGA and 45-nm ASIC evaluation},
journal = {Journal of Systems Architecture},
volume = {181},
pages = {104028},
year = {2026},
issn = {1383-7621},
doi = {https://doi.org/10.1016/j.sysarc.2026.104028},
url = {https://www.sciencedirect.com/science/article/pii/S1383762126003462},
author = {Dinis Santos and Joao Cabacinho and Joao Casaleiro and Valderi R.Q. Leithardt and Luis Bica Oliveira},

}

@INPROCEEDINGS{Sergio,
  author={Mendonça, Sérgio and Leithardt, Valderi R. Q. and Crocker, Paul A.},
  booktitle={2026 International Symposium on Networks, Computers and Communications (ISNCC)}, 
  title={Data Privacy and Security in the Quantum Era: A Comparative Evaluation between Classical and Post-Quantum Cryptography}, 
  year={2026},
  volume={},
  number={},
  pages={1-6},
  doi={10.1109/ISNCC70543.2026.11693868}}

@article{2026,
 title = {OntOraculum: a semantic ontology for managing performance metrics in smart environments},
 author = {Noetzold, Darlan and Leithardt, Valderi R. Q. and De Paz, Juan Francisco and Barbosa, Jorge L. V.},
 year = 2026,
 month = oct,
 volume = 12,
 pages = {e4078},
 journal = {PeerJ Computer Science},
 issn = {2376-5992},
 url = {https://doi.org/10.7717/peerj-cs.4078},
 doi = {10.7717/peerj-cs.4078}
}

\vspace{-3cm}

\vspace{-3cm}

\vspace{-3cm}

\end{document}